\documentclass[sigconf,authorversion,nonacm]{acmart}
\usepackage{xspace}
\usepackage{syntax}
\usepackage{listings}
\usepackage{xcolor}
\usepackage{colortbl}
\usepackage{textcomp}
\usepackage{multirow}
\usepackage{array}
\usepackage{stmaryrd}
\usepackage{amsmath}
\usepackage{subcaption}
\usepackage{booktabs}

\usepackage{enumitem}
\setlist[enumerate]{itemsep=0pt,topsep=0pt,parsep=0pt,partopsep=0pt}
\setlist[itemize]{itemsep=1pt,topsep=2pt,parsep=0pt,partopsep=0pt}
\setlist[description]{itemsep=0pt,topsep=0pt,parsep=0pt,partopsep=0pt}
\definecolor{codegreen}{rgb}{0,0.6,0}
\definecolor{codegray}{rgb}{0.5,0.5,0.5}
\definecolor{codepurple}{rgb}{0.58,0,0.82}
\definecolor{backcolour}{rgb}{0.95,0.95,0.92}
\lstdefinestyle{TAPstyle}{
	backgroundcolor=\color{backcolour},   
    commentstyle=\color{codegreen},
    keywordstyle=\color{magenta},
    numberstyle=\tiny\color{codegray},
    stringstyle=\color{codepurple},
    basicstyle=\ttfamily\footnotesize,
    breakatwhitespace=false,         
    breaklines=true,                 
    captionpos=b,                    
    keepspaces=true,                 
    numbers=left,                    
    numbersep=5pt,                  
    showspaces=false,                
    showstringspaces=false,
    showtabs=false,                  
    tabsize=2,
	morekeywords={rule,when,Item,received,then,end}
}
\renewcommand\appendix{\par
	\setcounter{section}{0}
	\setcounter{subsection}{0}
	\setcounter{figure}{0}
	\setcounter{table}{0}
	\setcounter{lstlisting}{0}
    \setcounter{theorem}{0}
	\renewcommand\thesection{\Alph{section}}
	\renewcommand\thefigure{\Alph{section}\arabic{figure}}
	\renewcommand\thetable{\Alph{section}\arabic{table}}
	\renewcommand\thelstlisting{\Alph{section}\arabic{lstlisting}}
	\renewcommand*{\theHsection}{\thesection}
	\renewcommand*{\theHsubsection}{\thesubsection}
	\renewcommand*{\theHlstlisting}{\thelstlisting}
}

\usepackage[normalem]{ulem}

\usepackage[T1,OT1]{fontenc} 

\newcommand{\etal}{\textit{et al.}\xspace}
\newcommand{\ie}{\textit{i.e.}\xspace}
\newcommand{\eg}{\textit{e.g.}\xspace}

\newcommand{\code}[1]{{\texttt{#1}}\xspace}

\newcommand{\mycomment}[1]{}
\newcommand{\FIGREF}{Figure\xspace}

\newtheorem{theorem}{Theorem}

\newcommand{\oh}{OpenHAB\xspace}
\newcommand{\st}{SmartThings\xspace}
\newcommand{\ifttt}{IFTTT\xspace}

\newcommand{\helion}{{H$\epsilon$lion}\xspace}
\newcommand{\sysname}{{{\textsc{Maverick}}}\xspace}
\newcommand{\approach}{\sysname}

\newcommand{\ra}[1]{\renewcommand{\arraystretch}{#1}}

\newcommand{\ISSP}{\ensuremath{\mathtt{ISSP}}\xspace}
\newcommand{\cmd}[1]{\ensuremath{\mathsf{#1}}\xspace}
\newcommand{\policy}{\ensuremath{\mathcal{P}}\xspace}

\newcommand{\cI}{\ensuremath{\mathcal{I}}\xspace}
\newcommand{\cS}{\ensuremath{\mathcal{S}}\xspace}
\newcommand{\cV}{\ensuremath{\mathcal{V}}\xspace}
\newcommand{\cD}{\ensuremath{\mathcal{D}}\xspace}
\newcommand{\cP}{\ensuremath{\mathbb{P}}\xspace}
\newcommand{\cA}{\ensuremath{\mathcal{A}}\xspace}
\newcommand{\cR}{\ensuremath{\mathcal{R}}\xspace}
\newcommand{\sAS}{\ensuremath{\mathcal{C}}\xspace}

\definecolor{Gray}{gray}{0.85}
\definecolor{LightCyan}{rgb}{0.88,1,1}

\newcolumntype{a}{>{\columncolor{Gray}}c}
\newcolumntype{b}{>{\columncolor{white}}c}

\newcommand{\notsupported}{{}\xspace}

\usepackage{tikz}
\newcommand*\circled[1]{\tikz[baseline=(char.base)]{
		\node[shape=circle,fill,inner sep=0.5pt] (char) {\bfseries\footnotesize\textcolor{white}{#1}};}}
	
\newcommand{\phyact}{\ensuremath{\cA_p}}
\newcommand{\cyberact}{\ensuremath{\cA_c}}

\begin{document}

%
\title{\sysname: An App-independent and Platform-agnostic Approach to 
Enforce Policies in IoT Systems at Runtime}


\author{M. Hammad Mazhar}
\orcid{0000-0001-8663-4343}
\affiliation{
	\department{Department of Computer Science}
	\institution{The University of Iowa}
	\city{Iowa City}
	\state{IA}
	\country{USA}
}
\email{muhammadhammad-mazhar@uiowa.edu}

\author{Li Li, Endadul Hoque}
\orcid{0009-0000-0437-6193}
\orcid{0000-0002-6682-9618}
\affiliation{
	\department{Department of EECS}
	\institution{Syracuse University}
	\city{Syracuse}
	\state{NY}
	\country{USA}
}
\email{{lli101,enhoque}@syr.edu}

\author{Omar Chowdhury}
\orcid{0000-0002-1356-6279}
\affiliation{
	\department{Department of Computer Science}
	\institution{Stony Brook University}
	\city{Stony Brook}
	\state{NY}
	\country{USA}
}
\email{omar@cs.stonybrook.edu}

\renewcommand{\shortauthors}{Mazhar, et al.}

\begin{abstract}
	%
	Many solutions have been proposed to curb unexpected behavior of automation apps installed on programmable IoT platforms by enforcing safety policies at runtime.
	However, all prior work addresses a weaker version of the actual problem due to a simpler, unrealistic threat model. 
	These solutions are not general enough as they are heavily dependent on the installed apps and catered to specific IoT platforms.
	%
	Here, we address a stronger version of the problem via a realistic threat model, where (i) undesired cyber actions 
	can come from not only automation platform backends (e.g., SmartThings) but also close-sourced third-party services (e.g., IFTTT), and (ii) physical actions (e.g., user interactions) on devices can move the IoT system to an undesirable state.
	We propose a runtime mechanism, dubbed \sysname, which employs an app-independent, platform-agnostic mediator to enforce policies against all undesired cyber actions and applies corrective-actions to bring the IoT system back to a safe state  from an unsafe state transition. 
	\sysname is equipped with a policy language capable of expressing rich temporal invariants and an automated toolchain that includes a policy synthesizer and a policy analyzer for user assistance. We implemented \sysname in a prototype and showed its efficacy in both physical and virtual testbeds, incurring minimal overhead.
\end{abstract}


\maketitle
\vspace*{-0.1in}
\section{Introduction}
\label{sec:intro}

Modern \textit{programmable IoT platforms} (\eg, \st, \oh, \ifttt) offer a cost-effective 
automation platform to install customized applications (aka, \textit{apps}), 
which in most cases require no explicit support from device (\eg, smart lock) manufacturers. 
Despite tremendous innovation, safety and security concerns in programmable IoT systems 
are still a pressing issue \cite{AttackSurface}. 
One source of this issue revolves around the automation itself. 
For instance, many apps can inconceivably interact with  installed devices leading to a 
disastrous outcome, which can compound due to a chain-reaction 
in a smart home with multiple apps. 
Many solutions have been proposed to curb unexpected behavior of apps installed on IoT platforms \cite{contexiot17,iotguard2019ndss,Ding2021IoTSafe,yahyazadeh2020patriot,yahyazadeh2019expat,celik2018soteria,barrera2018standardizing,chi2020crossapp,fernandes2016smart,kaffle2021homeendorser}.
One class of prior work hinges on the enforcement  policies at runtime \cite{contexiot17,Ding2021IoTSafe,iotguard2019ndss,yahyazadeh2019expat,yahyazadeh2020patriot,kaffle2021homeendorser} -- the focus of this paper.

\begin{figure}[!t]
	\centering
	\includegraphics[width=\columnwidth]{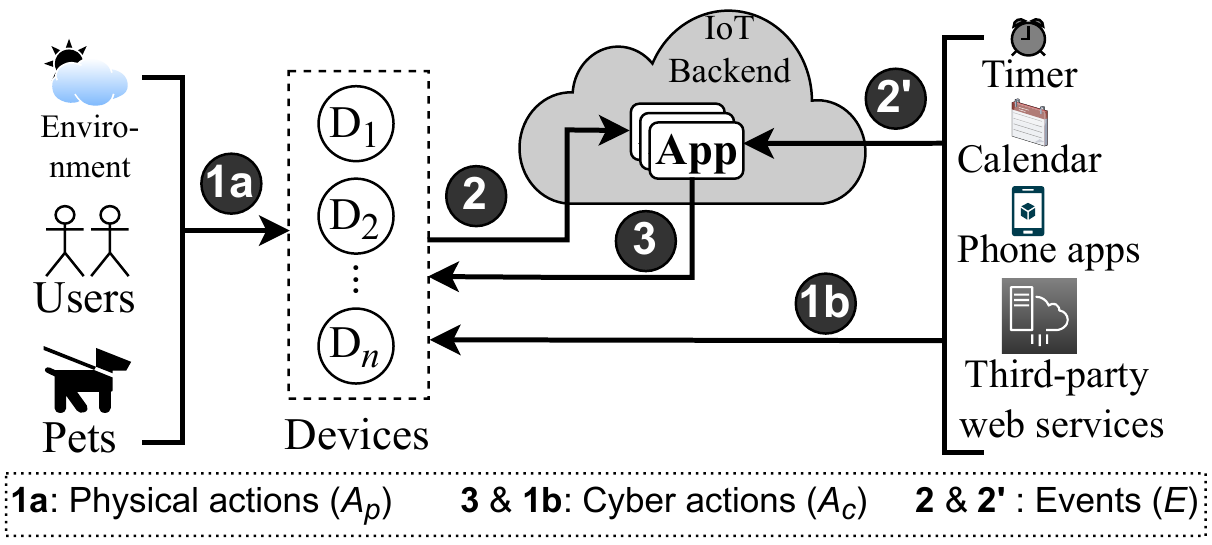}
	\vspace{-0.2in}
	\caption[A modern programmable IoT system.]{A modern programmable IoT system. \textbf{$D_1,...,D_n$} are IoT devices, \circled{1a} are physical interactions, \circled{1b} are cyber actions sent directly to devices, \circled{2} and \circled{2'} are events from IoT devices and from other services respectively, and \circled{3} are cyber actions generated by apps on an IoT backend.}	\label{fig:iot-sys}
	\vspace{-0.5in}
\end{figure}

For IoT systems,
a policy expresses what aspects of the system must not be violated by the current global state of the system
(aka, \textit{system-state}, composed of the current internal state of each installed device). 
An \textit{unsafe} state of the system is 
a system-state
that violates at least one policy at runtime.
During an unsafe state, the system's safety can be compromised, which can 
result in dire consequences. 
Any change in the system-state is due to a change in a device's internal state, which is in turn 
due to an \emph{action}/\emph{command} executed on the device (\eg, \code{lock/unlock}). Additionally, any change in a device's state
results in an \textit{event} (\eg, \code{locked/unlocked}), possibly triggering execution of one or more automation apps
(\circled{2} in \FIGREF~\ref{fig:iot-sys}).

Some actions can be undesired in the current system-state if they can induce a system 
transition into an unsafe state (\eg, unlocking the front door when nobody is at home). 
To ensure that the system obeys the given policies at runtime, one needs to prevent undesired actions 
from reaching the devices. 
\FIGREF~\ref{fig:iot-sys} shows the sources of undesired actions in a modern IoT system: 
\textit{physical actions} (\phyact) via user interactions or environmental changes (\circled{1a}) 
and \textit{cyber actions} (\cyberact) sent to devices through network messages (\circled{3} and \circled{1b}).

To the best of our knowledge, most prior work \cite{iotguard2019ndss,Ding2021IoTSafe,contexiot17,yahyazadeh2020patriot,yahyazadeh2019expat}
assumes a weaker threat model that considers
 apps installed in the IoT platform's native backend (\circled{3} in \FIGREF~\ref{fig:iot-sys})
as the only source of undesired actions.
Technically, they focus on solving whether installed apps violate given policies -- 
a \textit{weaker version of the actual problem}. 
These proposals are heavily app-dependent 
(\eg, require source code)
and thus platform-specific. Most safely assume \circled{1a} and \circled{1b} actions 
can never transition the system into an unsafe state, which is \textit{unrealistic in practice}.
If the system ever transitions into an unsafe state 
due to 
\circled{1a} or \circled{1b}, these solutions will either be stuck in limbo \cite{yahyazadeh2019expat,yahyazadeh2020patriot} or 
inaccurately enforce policies \cite{iotguard2019ndss}.%


In this paper, we tackle the \textit{stronger version of the actual problem} where we want to ensure 
that the deployed IoT system satisfies the user-specified policies at runtime, 
independent of IoT backend apps and third party services.
We expand the threat model where the sources of undesired actions include 
all possible cyber and physical actions (\circled{1a}, \circled{1b}, and \circled{3} in \FIGREF~\ref{fig:iot-sys}). 
Using this threat model, we \textit{formally} define the stronger version of the problem
as the \textit{IoT System Safety Problem} (\ISSP) for a general IoT system.
\ISSP demands a runtime mechanism 
that ensures that (\textbf{R1}) a cyber action that transitions the IoT system into an unsafe state is prevented from the reaching the target device, 
and (\textbf{R2}) if the system ever transitions into an unsafe state due to a physical action 
(or a change in the environment), the mechanism can take appropriate actions 
to bring the system back to a safe state.
We propose a runtime approach, dubbed \textbf{\sysname}, 
that ensures both \textbf{R1} and \textbf{R2}, while being independent of installed apps 
and IoT platforms. 


To ensure \textbf{R1} under limited access to apps' source code, 
a solution must be able to mediate all cyber actions (\circled{1b} and \circled{3}). 
\sysname tackles this issue by employing a \textbf{platform-agnostic mediator} which 
mediates all communication between apps and IoT devices (\ie, \circled{2}, \circled{3}, and \circled{1b} in 
\FIGREF~\ref{fig:iot-sys}) and enforces user-specified policies at runtime ($\S$~\ref{sec:overview}).
Unlike prior policy enforcement work, we deliberately design this mediator to operate as an individual component (similar to PFirewall\cite{chi2021pfirewall})
 independent of the platform's backend and app engine, 
third-party services, and installed apps.
The mediator 
is equipped with the necessary communication interfaces to act as a man-in-the-middle, but \textit{trusted intermediary},
that intercepts all messages between IoT devices and its backend or third-party services. 
For reliable runtime policy enforcement, the mediator also maintains a \textit{shadow} system-state
 in-sync with IoT device state by monitoring state update messages. 
For \textbf{R2}, \sysname redefines the concept of \textbf{corrective actions}. 
A corrective action is technically a cyber action to be sent to an IoT device or a drop of a violating cyber action. Whenever \sysname finds 
that the IoT system is in an unsafe state and violates a policy (say $P$), \sysname applies a sequence of corrective actions carefully chosen for $P$
 so that the sequence 
can nudge the system toward a safe state that satisfies $P$. 

Often policies for a modern IoT system cannot be enforced reliably without knowledge of system execution history
capturing temporal ordering of events and actions. Some policies are also contingent on environment/system variables.
Unlike prior work, we develop a \textit{policy language} ($\S$~\ref{sec:policy-lang}) that can express conditions on temporal ordering 
of past events/actions,
rich attributes of IoT devices and their relationships, and custom predicates on system variables.
\sysname expects the user to specify the desired policies in this language. 
For each policy, the user also specifies the corresponding corrective actions.

Asking users to write policy in a dedicated policy language 
can impede the adoption of \sysname in practice. 
To address this,
we augment our policy language toolchain with two automated components for policy synthesis ($\S$~\ref{sec:policy-authoring}):  
\textit{invariant synthesizer} and \textit{policy analyzer}.
The invariant synthesizer aims to lift the burden of the user by generating invariants 
(\ie, fine-grained components of a policy)
based on a user-provided set of positive and negative execution traces of the system
such that the positive traces satisfy the invariants but the negative ones violate them. 
The policy analyzer checks the effectiveness of user-selected corrective actions.

We also prove that the decision version of \ISSP 
is \textit{undecidable} in general.
In other words, once the system transitions to an unsafe state, 
whether we can nudge the system back to a safe state using corrective actions
is undecidable. 
Despite this undecidability, it is possible to develop a \textit{sound} solution 
for the ISSP decision problem under reasonable assumptions ($\S$~\ref{sec:policy-lang}).
To evaluate if user-provided corrective actions are sufficient to bring the system back
to a safe state, we utilize our policy analyzer, reducing this problem 
to a model checking problem ($\S$~\ref{sec:policy-authoring}).  

We implement a prototype of \sysname on top of a Mosquito MQTT server \cite{mosquitto}, which 
communicates with different IoT platform backends and third-party services.
Our prototype leverages \oh to enable support 
for numerous WPAN technologies (\eg, Z-wave, ZigBee) for IoT device communication.
The mediator of our prototype translates messages  between the MQTT network and the WPAN (\eg, ZigBee)
and enforces policies or applies corrective actions as required at runtime.
We have empirically evaluated our prototype in both virtual and physical testbeds with 
system execution examples/scenarios drawn from an existing dataset \cite{manandhar2020Helion}.
Our evaluation reports that \sysname is highly effective while incurring minimal overhead on latency and throughput.

\noindent\textbf{Contributions.} This paper has the following contributions: 
\begin{enumerate}[wide,labelwidth=!,labelindent=0pt]
	\item We introduce the IoT system safety problem (\ISSP) which captures a realistic threat model.	
	We show that the decision problem version of \ISSP is undecidable in general; however, 
	we develop a sound algorithm under some assumptions. 
	
	\item We propose a runtime solution, dubbed \sysname, which enforces policies against undesired cyber actions from 
	different sources by employing a {platform-agnostic trusted mediator} and
	applies {corrective actions} to recover the IoT system from an unsafe state. 
	
	\item 
	To assist users write policies, we develop a policy language to capture temporal invariants 
	and its necessary automated toolchain, including a policy/invariant synthesizer and a policy analyzer. 
	 
%
	\item We implemented the features of \approach and evaluated all its components 
	on both virtual and physical test-beds using an existing dataset. 
	Our evaluation shows that \approach is effective and incurs a minimal overhead.
\end{enumerate}

\section{Preliminaries}
\label{sec:preliminary}
\noindent\textbf{IoT Devices.} 
\emph{IoT devices} are the main component of an IoT system. 
IoT device functionality can be roughly partitioned into 
two categories: \emph{sensing} and \emph{actuation}. An IoT device uses its sensing capabilities to discern one or more environmental 
events (e.g., change of temperature, presence). 
An IoT device uses its \emph{actuation} capabilities to carry out some tasks based 
on some \emph{cyber} actions (\cyberact) or \emph{physical} actions (\phyact) (e.g., sending an \cmd{on} command 
or pressing the physical switch). An IoT device may be equipped with 
multiple sensing and actuation functionality, and also maintains one or more 
states based on such functionality. 
An IoT device takes one or more events (e.g., environmental 
change (\circled{1a}), or a cyber/physical command (\circled{1a},\circled{1b} or \circled{3})) as inputs, changes its internal state, 
and finally generates an output event (\circled{2}) confirming the state update. 

\noindent\textbf{IoT Hubs/Gateways.}
An IoT device may communicate over the Internet directly through Wi-Fi or a wired connection, or utilize WPAN technologies such as as Z-Wave \cite{zwave}, Zigbee \cite{zigbee}, Bluetooth BLE \cite{gupta2016inside}.
WPAN connectivity is facilitated by \textit{device hubs} which then provide IP-based connectivity with a device-specific mobile \textit{companion app} or with \textit{automation platforms}.
These automation platforms may be hosted on \textit{local gateways} placed within the home network (e.g. OpenHAB), or hosted on cloud backends and placed remotely (e.g Samsung SmartThings, IFTTT).
Cloud-hosted automation platforms may either communicate to the IoT device itself directly or via its device hub, or utilize a local gateway. 

\noindent\textbf{Automation platforms.}
These allow users to automate the behavior of their IoT devices through the use of \emph{apps}.
Automation services provide a centralized interface to monitor and control behavior of devices from different manufacturers, utilizing manufacturer-provided integrations. 
The automation services can also be categorized into \emph{native} and \emph{third-party} services. In the former, 
users have fine-grained control over how the installed apps are executed. For example, one can rewrite an installed app to include 
policy checking hooks, such as in \oh and Samsung \st. 
Third-party automation services provide a limited programming interface and control over app executions. In those platforms (e.g., 
IFTTT, Zapier), users can only choose the trigger conditions and the corresponding commands to the IoT devices. 
Apps on those 
platforms cannot be equipped with custom policy checking hooks and hence are not amenable to existing runtime approaches \cite{contexiot17,iotguard2019ndss,yahyazadeh2019expat,yahyazadeh2020patriot}.

\noindent\textbf{Programmable IoT System.} 
One can view an IoT system as a distributed system with IoT devices and hubs/gateways as individual computational components. Many automation platforms (e.g., IFTTT) 
provide programming interfaces for customizing system operation without 
considering low-level implementation details (e.g., event serialization). An IoT system integrated with such platforms is called a \emph{programmable IoT system}, as illustrated in Figure \ref{fig:iot-sys}. 
It primarily provides automation to orchestrate IoT devices 
to perform a set of high-level, complex actions when certain conditions are fulfilled, \eg automatically opening the 
curtains, brewing coffee, and starting the toaster when the user wakes up in the morning (See Listing \ref{code:TAP} in the appendix). The underlying automation 
unit of these automation services is an \emph{automation application/app/rule} which can be viewed as a (stateful) 
event-driven program and often takes the form of a \emph{trigger action rule}. 
Other platforms also support full-fledged event-driven programs, with sandboxed automation app execution.
When an event 
reaches the automation platform, it feeds that event to all automation apps registered for the trigger event and then executes them. Any actions generated by the automation apps (e.g., giving the toaster the \cmd{switch\_on}  command) are  routed back to the appropriate IoT devices. 


\vspace*{-0.15in}
\section{Motivation and Problem Statement}\label{sec:problem}
We now formally define the programmable IoT system safety problem (\ISSP)
along with its complexity. 
We start the section by discussing 
the threat model, a running example, and limitations 
of current work in solving \ISSP in its 
entirety. 

\noindent\textbf{Threat Model.} 
We consider that a user can install automation apps in various automation services, including, 
native (\eg, \st) and third-party (\eg, \ifttt) platforms. 
The source code for such apps are not always accessible or modifiable. 
These apps can push the IoT system to transition to a unsafe state where some desired policies are violated
(\eg, leaving the oven on when the user is away, opening the front door when the user is sleeping).  
While most prior work \cite{iotguard2019ndss,celik2018soteria,yahyazadeh2020patriot,yahyazadeh2019expat,contexiot17} considers the apps -- installed in the native platform (\circled{3} in \FIGREF~\ref{fig:iot-sys}) -- as the source of
undesired actions, we consider a stronger version where the source of undesired actions
consists of native apps \circled{3}, third-party apps \circled{1b}, and physical interactions \circled{1a}.
We assume that devices are tamper-proof, report state faithfully and respond faithfully to given commands.
Sensor-spoofing and network-level attacks are outside the scope.

With our threat model, prior work \cite{celik2018soteria,yahyazadeh2019expat,yahyazadeh2020patriot,contexiot17} is largely ineffective, because they only focus on \circled{3} and 
rely on apps' source code instrumentation to place hooks for collecting information and enforcing policies.
While Expat \cite{yahyazadeh2019expat} and Patriot \cite{yahyazadeh2020patriot} make the policy decision inside the instrumented apps, 
IoTGuard \cite{iotguard2019ndss} and IoTSafe \cite{Ding2021IoTSafe} employ 
a remote server for their policy decision due to the lack of access (\ie, closed-source-ness) of their chosen platform
(\eg, SmartThings). 
Third-party apps (\eg, \ifttt) are mostly closed-source with limited user instrumentation opportunities. 
One can argue that \ifttt provides its users with much more control of the execution of its apps, which 
can be leveraged to enforce policies directly inside the apps, as in prior work for \circled{3} 
through ``filter code'' that modifies actions or aborts executions \cite{IFTTTfiltercode}.
However, it is not the same as
intercepting and blocking an undesired action by an external policy enforcer at runtime.
While the former needs to be in-built in the source, the latter should be performed
outside the \ifttt server at runtime, unbeknownst to \ifttt. 

Physical interactions \circled{1a} with devices cannot be intercepted by any IoT apps/platforms. It is possible that
the system can transition to an unsafe state due a physical interaction with a device 
(\eg, the user turns off the water valve). 
Prior work considers no such physical interactions, and hence, if the system ever
goes to an unsafe state, their solution does not guarantee any runtime defense. For instance, Expat \cite{yahyazadeh2019expat},
Patriot \cite{yahyazadeh2020patriot}, IoTGuard \cite{iotguard2019ndss} will inaccurately allow/block actions if the system is in an unsafe state.

All policies we consider actually capture the user's expectation from the system. 
A policy violation indicates that there exists a combined execution 
of the installed apps (possibly, collected from unvetted sources) 
with or without physical interactions such that it leads the system to behave unexpectedly, 
which technically violates the safety of the system. 
In that sense, all our policies essentially evaluate the safety of the system and any violation 
that impacts it.


\noindent\textbf{Motivating example.} 
To demonstrate limitations in prior work, we use an example from existing work \cite{iotguard2019ndss} with a slight 
deviation.
Consider a programmable smart home with a water leakage sensor, a smart water system 
whose main valve can be switched off using a command, a smart sprinkler system \cite{ssprinkler}, 
a temperature sensor, and a smoke sensor. 
Assume the user installs two apps, namely, \textit{fire protection app} (FPA) and 
\textit{water-leakage protection app} (WPA). 
The FPA switches on the smart sprinkler system when the smoke sensor senses smoke 
and the temperature is higher than some threshold suggesting potential fire. 
The WPA switches off the main water valve whenever the water leakage sensor senses flooding. 
In case of a fire, the smart sprinkler system will go off due to FPA. 
After sometime, however, the water leak sensor will sense flooding, 
and WPA will switch off the main water valve, cutting water to the sprinklers and 
letting the fire go on unchecked. 
The ideal response here is to ignore the flooding and douse the fire first, by preventing WPA from switching off the main water valve.

While some prior work \cite{yahyazadeh2019expat, yahyazadeh2020patriot}
can take the ideal measure with 
appropriate policy, IoTGuard \cite{iotguard2019ndss} cannot take the ideal 
measure even with proper policy as its policy enforcement mechanism is based on a reachability analysis,
which cannot handle app interactions through physical channels (\eg,  air) \cite{Ding2021IoTSafe}. 
We learned this counter-intuitive fact from an attempt
to implement it 
\footnote{The full implementation of IoTGuard \cite{iotguard2019ndss} is not publicly available}.

Now we extend our example to cover all types of cyber actions (\circled{3} and \circled{1b} in \FIGREF~\ref{fig:iot-sys}).
Suppose the user installed FPA in 
the smart home's native IoT backend 
and installed WPA in an third-party service (\eg, \ifttt) as in this app \cite{wpa},
where the user does not 
access to the WPA's actual source code.%
\footnote{{Installing automation apps in a third-party service is a reasonable assumption with device-specific companion phone apps which can set up such automation. Unlike FPA, WPA will be operated, controlled and monitored through the third-party’s own cloud-based backend.}} 
While cyber actions sent by FPA represents \circled{3}, 
 cyber actions by WPA represents \circled{1b}.
Most prior work \cite{yahyazadeh2019expat,yahyazadeh2020patriot,Ding2021IoTSafe,iotguard2019ndss} 
cannot prevent WPA from switching off the main water valve, because of their 
inability to mediate cyber actions from apps running on a third-party service.%
\footnote{Prior work like IoTGuard \cite{iotguard2019ndss} and others \cite{yahyazadeh2019expat,yahyazadeh2020patriot,Ding2021IoTSafe} cannot actually 
mediate the cyber actions sent by \ifttt apps, because they 
rely on the instrumentation of apps' source code to mediate actions and enforce policies,
but unfortunately an \ifttt app's source code is not available and thus cannot be instrumented.
}
%

Next we incorporate some physical actions (\circled{1a} in \FIGREF~\ref{fig:iot-sys}) in our example. 
Assume the user has turned off the main water value physically (perhaps for maintenance)
and forgot to turn it back on; as a result, the system has moved to an unsafe state. Even if FPA turns on the sprinklers, no water will be sprayed because the main water valve is closed.
Most of prior work \cite{yahyazadeh2019expat,yahyazadeh2020patriot,Ding2021IoTSafe,iotguard2019ndss} cannot 
take any measure to recover the IoT system from this unsafe state. 

\begin{table*}[!ht]
	\newcounter{featno}
	\renewcommand{\thefeatno}{\arabic{featno}}
	\newcommand{\feat}[1]{%
		\refstepcounter{featno}
		\label{#1}
		F\thefeatno\xspace%
	}
	
	\small
	\centering
	 \ra{1.1}
	\footnotesize
	\begin{tabular}{@{}l|l|c|c|c|c|c|c@{}}
		\hline
		\textbf{F\#} & \textbf{Feature}                                                           &   IotGuard \cite{iotguard2019ndss}    &     ExPAT \cite{yahyazadeh2019expat}     &    PatrIoT \cite{yahyazadeh2020patriot}    &   ContexIoT \cite{contexiot17}   &    IoTSafe \cite{Ding2021IoTSafe}     &   \approach   \\ \hline\hline
		\feat{feat:threat}                & Stronger threat model                                                      & \notsupported & \notsupported & \notsupported & \notsupported & \notsupported  &  \checkmark   \\ \hline
		\feat{feat:cyber3}                & Mediate cyber actions \circled{3} in \FIGREF~\ref{fig:iot-sys}             &  \checkmark   &  \checkmark   &  \checkmark   &  \checkmark   &   \checkmark   &  \checkmark   \\ \hline
		\feat{feat:cyber1b}               & Mediate cyber actions \circled{1b} in \FIGREF~\ref{fig:iot-sys}            & \notsupported & \notsupported & \notsupported & \notsupported & \notsupported  &  \checkmark   \\ \hline
		\feat{feat:recovery}              & Recover from unsafe state due to \circled{1a} in \FIGREF~\ref{fig:iot-sys} & \notsupported & \notsupported & \notsupported & \notsupported & $\circleddash$ &  \checkmark   \\ \hline
		\feat{feat:decouple}              & Clear distinction between apps and policies                                &               &  \checkmark   &  \checkmark   &               &                &  \checkmark   \\ \hline
		\hline
		\feat{feat:appdep}                & App-dependent core mechanism                                               &  \checkmark   &               &               &  \checkmark   &   \checkmark   & \notsupported \\ \hline
		\feat{feat:instrument}            & Require app instrumentation                                                &  \checkmark   &  \checkmark   &  \checkmark   &  \checkmark   &   \checkmark   & \notsupported \\ \hline
		\feat{feat:outsync}               & Out-of-sync state management                                               &  \checkmark   &               &               &      \checkmark       &   \checkmark   & \notsupported \\ \hline
		\feat{feat:manual}                & Require user involvement during policy enforcement                         &  \checkmark   &               &               &  \checkmark   &                & \notsupported \\ \hline
	\end{tabular}
	\vspace{-0.15in}
	\caption{Comparison of \approach with IotGuard \cite{iotguard2019ndss}, EXPAT \cite{yahyazadeh2019expat}, PatrIoT \cite{yahyazadeh2020patriot}, ContexIoT \cite{contexiot17}, and IoTSafe \cite{Ding2021IoTSafe}. \checkmark refers to the feature being supported/available/implemented whereas $\circleddash$ refers to the feature being partially supported/implemented 
		and an empty cell refers to the feature being not required/implemented/supported/applicable.}
	\label{tab:capabilities}
	\vspace{-0.3in}
\end{table*}

\noindent\textbf{Inadequacy of prior work.}
Table~\ref{tab:capabilities} outlines how prior work 
\cite{iotguard2019ndss, yahyazadeh2019expat, yahyazadeh2020patriot, contexiot17, Ding2021IoTSafe}
closely related to \sysname are quite inadequate in terms of multiple perspectives, 
highly important to the IoT safety problem.
Here we only concentrate on some key perspectives, 
and the rest will be covered later in the paper. 

No prior work supports the stronger threat model (F\ref{feat:threat}) that \sysname considers to address as outlined earlier. 
Unlike \sysname, the need to instrument apps' source code (F\ref{feat:instrument}) impedes 
their mediation of cyber actions \circled{1b} sent by the apps running on third-party services (F\ref{feat:cyber1b}) 
where the access to an app's source code is not allowed.
Much of existing work is heavily dependent and fully driven by the installed apps (F\ref{feat:appdep}). For instance, 
IoTGuard \cite{iotguard2019ndss} and IoTSafe \cite{Ding2021IoTSafe} rely on app interactions to generate dynamic interaction graphs, used in their policy enforcement step at runtime.
These graphs record and manage a shadow copy of the system-state for their outsourced policy checker. 
A change in a device's state (aka, an event) that does not trigger any installed app, will never be 
recorded in the generated graph, resulting in a shadow system-state that is out-of-sync with the reality (F\ref{feat:outsync}), leading to inaccurate and unreliable policy enforcement even during safe states.


Automation apps dictate what actions are taken when a specific set of events occurs, while
policies express invariants the system must not violate. 
Barring Expat \cite{yahyazadeh2019expat} and Patriot \cite{yahyazadeh2020patriot}, 
existing works (\eg, IoTGuard \cite{iotguard2019ndss}, IoTSafe \cite{Ding2021IoTSafe}, ContexIoT \cite{contexiot17}) cannot establish a clear distinction between apps and policies (F\ref{feat:decouple}), and their implemented policy enforcement mechanisms become incomprehensible in many of their own test cases.  
However, it appears that some of IoTSafe's \cite{Ding2021IoTSafe} policies can \textit{inadvertently} recover the 
system from an unsafe state caused by \circled{1a} actions (F\ref{feat:recovery}). 
Further investigation reveals that those policies are much like apps initiating 
some cyber actions (akin to our corrective actions)
rather than policies defining invariants.
As IoTGuard's \cite{iotguard2019ndss} policies can be much more general, policy enforcement can involve the user (F\ref{feat:manual}) to actively 
resolve conflicts, while ContexIoT \cite{contexiot17} always defers to the user 
when an app contemplates an action of interest.

\noindent\textbf{IoT System Safety Problem (\ISSP).} A given programmable IoT system \cI can be viewed as a 
state machine and can be expressed as a tuple $\langle \cS, \cV, \cD, \cA, \cP, \cR\rangle$. 
\cV is the (possibly, infinite) set of variables whose values can be drawn from the (possibly, 
infinite) set of constants \cD while making sure the types of variables are respected (e.g., a 
variable of type real can be assigned only real values, not string constants). \cS is the 
(possibly, infinite) set of states where each of which $s\in\cS$ maps all variables $v\in\cV$ to 
a value from $\cD$. Suppose $\cV=\{\mathsf{BedroomLightState},\ldots\}$ and $\cD=\{\mathit{lightON},\mathit{lightOFF},\ldots\}$, 
then a possible state $s\in\cS$ could be $s=[\mathsf{BedroomLightState}\mapsto\mathit{lightON}, \ldots]$. 
\cA is a finite set of high level commands that can be issued to the devices and can be decomposed into 
cyber commands (i.e., $\cA_c$) to be issued by automation services and human commands to be issued by the human or environment (i.e., $\cA_e$) where 
$\cA = \cA_c \cup \cA_e$.  
$\cP: \cS \rightarrow \{\mathsf{true}, \mathsf{false}\}\times \sAS$ is a function that takes a state and decides whether 
that state is safe or not. If the state is not safe, it also returns a sequence of \emph{corrective actions} $C\in\sAS$, each action in $C$ 
is drawn from the set $\cA_c$. $\cR\subseteq \cS\times \cA \times \cS$ is the left-total transition relation, which 
given the current state $s\in\cS$ and a contemplated action $a\in\cA$, decides which state \cI moves to next. 
\emph{The \ISSP problem requires that \cI never moves to an unsafe state defined by a given policy \policy, and even if it moves to an unsafe state, possibly 
due to an action in $\cA_e$, then it eventually moves back to a safe state according to the given policy \policy}. As we show below, the decision 
version of \ISSP is undecidable. 
\begin{theorem}[Decision \ISSP]
	Given a programmable IoT system $I=\langle \cS, \cV, \cD, \cA, \policy, R\rangle$ where $\policy\in\cP$ and 
	an unsafe state $s\in\cS$, deciding whether $I$ can be transitioned to a safe state $s^\star\in\cS$ according 
	to \policy is an undecidable problem.    
\end{theorem}
The proof reduces the Turing machine halting problem to the decision version of \ISSP (See Appendix \ref{appendix:isspProof}). 

\vspace*{-0.05in}
\section{\sysname Overview}\label{sec:overview}
We now provide a design overview of \sysname, highlighting each component.
As discussed in Section \ref{sec:preliminary}, it is possible 
to organize IoT devices, and different automation services and hubs 
into a variety of IoT system architectures.
Providing policy checking in these different IoT system architectures in 
an app- and platform-agnostic
way makes designing \approach challenging.
The main design decision is where to place the \approach's policy checker in an IoT system such that it has a full view of all the cyber actions 
and state updates without requiring explicit support from device manufacturers or automation service providers.
%
\begin{figure}[htbp]
    \centering
    \includegraphics[width=\columnwidth]{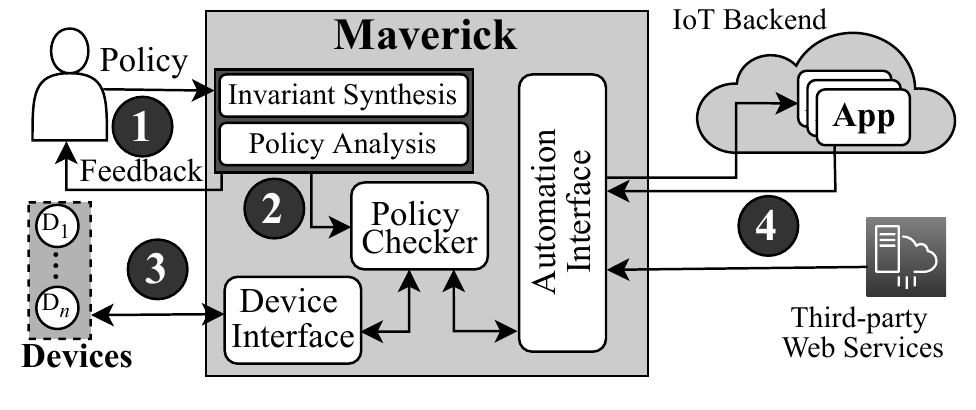}
    \caption{Overview of \sysname's design}
    \label{fig:internals}
    \vspace*{-0.3in}
\end{figure}
We design \approach as a man-in-the-middle, but trusted intermediary, 
illustrated in Figure \ref{fig:internals}.
\approach can be viewed as an amalgamation of the following three components: 
(a) \emph{A local device interface} to which devices connect using a supported communication 
protocol (e.g., ZigBee, Z-Wave, etc.) and share status updates (aka, events). \approach 
issues commands (predominantly, relaying commands from automation apps) to connected 
devices using this interface. 
(b) \emph{An automation service interface} to which the automation services connect for issuing commands to 
installed IoT devices. The same interface is used by \approach to forward any events to
automation services. 
(c) \emph{A policy checker}, which receives events from IoT devices and commands from different services and checks for safety violation. In case of safety violation, it issues relevant corrective actions. It also has an internal 
timer which triggers the policy/invariant checking.
The policy checker also maintains the \emph{shadow system state} through  events it receives from IoT devices.

\noindent\textbf{Workflow.} 
The user first connects their IoT devices to  \sysname's local device 
interface and registers all automation services (including, IoT platform backends and third-party services)
using \sysname's automation service interface. 
The user then writes policies 
by interacting with its \emph{invariant synthesizer} and \emph{policy analyzer} (\circled{1}), which aids users 
in writing policies in \approach{}'s policy language. The policy is then handed to 
the policy checker (\circled{2}).

Status updates from devices are forwarded 
to \approach{}'s device interface (\circled{3}), which then forwards them to the policy checker. If all policy invariants are 
satisfied, then device updates are forwarded to registered automation services and the shadow state is updated. If any invariants are violated, the policy checker issues corrective actions via the device interface and updates the shadow state from resulting device state updates. 
When automation services issue commands for devices through \approach{}'s automation 
service interface (\circled{4}), the policy checker is consulted. If all invariants will be satisfied upon acting on a command, 
it is forwarded to the target device and the shadow state is updated upon receiving device state updates. Otherwise, commands are dropped. 

Any device that can connect with a hub using Bluetooth/BLE, ZigBee, Z-wave, 
MQTT, and Wi-Fi is supported by \approach. Connecting a device or automation service to the \sysname does not require any modifications 
to the device or automation service. 
However, devices that connect to automation services through SSL/TLS-protected IP connections are not supported. 


\vspace*{-0.1in}
\section{Policy Language and Authoring}\label{sec:policy-lang}
We now describe the policy language and the associated tools we provide to help users write policies for \approach. 
A policy in \approach is not only used to define whether a given IoT system state is \emph{\textbf{safe}} but also 
used to provide corrective actions that can nudge the system to move towards a safe state from an unsafe state. 
\vspace*{-0.25in}
\subsection{Policy Language and Policy Checking}
\label{sec:policy-checking}
A \emph{policy} $\policy\in\cP$ in \approach is 
a sequence of \emph{policy rules} $\wp_j$ where $j\in\mathbb{N}$. Each policy rule 
$\wp_j$ is of the form $\langle I_j : C_j\rangle$ in which $I_j$ denotes the \emph{invariant} that 
must always be maintained by the IoT system to be safe whereas $C_j$ is the \emph{sequence of corrective actions} that must be taken if $I_j$ 
were to be violated. Any state $s\in\cS$ of the programmable IoT system \cI is considered \emph{\textbf{unsafe}} if and only if it violates 
at least one of the invariants $I_j$. 
%
The corrective actions component of a policy rule $\wp_j$, $C_j$, is a sequence and has the form $[a_0, a_1, a_2, \ldots, a_{m-1}]$ in which 
each $a_i\in\cA_c$ is a command to an IoT device to change its state. The available commands $a_i$ that can be used in the $C_j$ component 
of a policy rule depends on the installed IoT devices. 
The abstract syntax of an invariant $I_j$ is as follows. 
\vspace*{-0.05in}  
\begin{align*}
\langle\mathsf{Invariant }\rangle \quad  I &::=     \mathbf{If} \  \Phi \  \mathbf{Then} \  \Psi\\
\langle\mathsf{IfCondition}\rangle\quad  \Phi &::= p\mid \Phi_1\ \mathbf{and}\ \Phi_2\mid\mathbf{not}\  \Phi\mid\textbf{true}\\ 
\langle\mathsf{ThenCondition}\rangle\quad  \Psi &::= \Phi \mid \Psi_1\ \mathbf{since}\ \Psi_2\mid\mathbf{yesterday}\  \Psi\\
\langle\mathsf{Predicate}\rangle\quad  p &::=  t_1 \boxplus t_2 \mid  p_1\ \mathbf{and}\  p_2\mid\mathbf{not}\  p\mid\textbf{true}\\
\langle\mathsf{Term}\rangle\quad  t &::= x\in\mathcal{V} \mid \langle\mathsf{const.}\rangle  \mid \mathsf{func.}(t_1, \ldots, t_n)
\end{align*} 
\vspace*{-0.15in}  

An invariant $I$ takes the form:  
``\textbf{If} $\Phi$ is true \textbf{Then} $\Psi$ must be true''. In an invariant $I \equiv \mathbf{If} \  \Phi \  \mathbf{Then} \  \Psi$, 
$\Phi$ can be a predicate $p$, the constant true, or their logical combinations. In addition to the form of $\Phi$, 
the then-condition  $\Psi$  also allows standard past temporal operators (i.e., since and yesterday)  found in linear temporal logic (LTL) \cite{ltl}. This allows for richer invariant expression through consideration of past executed events.  
One of the atomic components of the invariant language is a predicate $p$. A predicate can be a relational operator $\boxplus$ (e.g., $\leq, \neq, $)
applied to a pair of terms or logical combinations of multiple predicates. A term is a variable $x$ drawn from the set of variables $\mathcal{V}$, 
a constant (e.g., ``ON'', 2, 4.9), or a function applied to one or more terms (e.g., $\mathsf{RoomTemperature}+10$).

\noindent\emph{\textbf{Example.}}
    A user wishes to devise a policy that ensures that their front door is locked whenever they are away from home.
    The invariant for this policy can be defined as ``$\mathbf{If}\  \mathsf{Away}\  \mathbf{Then} \ \mathsf{FrontDoorLocked}$'' with predicates defined as:
    \begin{align*}
    \mathsf{Away} & = (\text{HomeMode.status == ``Away''})\\[-0.5ex]
    \mathsf{FrontDoorLocked} & = (\text{FrontDoorLock.status == ``locked''})
    \vspace*{-0.1in}
    \end{align*}
    
    This invariant can be violated in two ways:
    (1) a command to ``unlock'' the door received from the automation system is allowed to reach the device or (2) the front door lock reports an ``unlocked'' status when the home mode is reported as away.
    For (1), the corrective action is to drop the ``unlock'' command and for (2), the front door is relocked by sending a ``lock'' command to it. 
    This set of corrective actions is defined as $C=\{\mathsf{\text{drop(``unlock'')}}, 
        \mathsf{\text{(send,`lock'')}}\}$    
    %
    The resultant policy is defined as $\policy = [\langle I,\ C \rangle]$.


\noindent\textbf{Policy checking.} \approach{}'s policy checker gets triggered by one of three conditions: (1) a downstream command is being sent to a 
device (\circled{1b} and \circled{3} in Figure \ref{fig:iot-sys}); (2) a device changed its state and a corresponding state update message is being sent upstream (\circled{2} in Figure \ref{fig:iot-sys}); (3) the global policy checker timer expired. 
In all cases, the policy checker checks 
each invariant for violation. If a violation is found, the policy checker triggers the associated corrective actions. The corrective actions are, 
however, not regulated by the policy checker. The policy checking in condition (1) ensures that \sysname covers all possible sources of cyber actions as compared to prior work which depends upon triggering of an instrumented app or platform API call, 
\approach can preemptively prevent the IoT system from transitioning into an unsafe state 
 by denying the violating commands (i.e., through the special ``$\mathsf{drop}$'' corrective action).
The policy checking for condition (2) is necessary for bringing the system to safe state in the case the system transitioned to an unsafe state 
due to a physical action or a command issued to the device 
through a channel that is unobservable by \approach.   Policy checking in condition (3) is needed when it takes multiple iterations of 
corrective actions to navigate the system to a safe state. 

\approach{}'s policy checker takes as input the current state $\sigma_i$, the execution history $\sigma$, and the policy \policy, then it generates a 
sequence of corrective actions (possibly, empty if all invariants hold). In addition, for condition (1), it also allows the contemplated 
downstream command to reach the target device. Each invariant in our policy can be 
represented in a restricted, first-order linear temporal logic formula (i.e., no quantifiers). One can use dynamic programming based approach from the runtime 
verification literature \cite{monpoly,echeverriaphoenix} to check the invariant in an efficient way without having to store the whole execution history of the system (See Appendix \ref{appendix:policyChecking} for details). 
 
\vspace*{-0.15in}
\subsection{Policy Synthesis and Analysis}
\label{sec:policy-authoring}

We now discuss tools we have designed to help 
users  write policies. 

\noindent\textbf{Policy Synthesis Workflow.} The policy 
synthesis workflow (\circled{1} in Figure \ref{fig:internals}) starts by the user identifying 
a set of scenarios that are undesirable (i.e., unsafe) 
for the IoT system. For each scenario, 
the user generates one or more diverse, example system executions/\emph{traces} of changes in IoT device states 
in which the undesired state did not arise (i.e., \emph{positive examples}). The users then generate a diverse set of example system execution where the 
undesired state occurred (i.e., \emph{negative examples}). Such traces may vary in terms of event count or precision of the relevant state. The user feeds those example traces 
into the \emph{\textbf{automatic invariant synthesizer}}, which outputs a set of \emph{candidate invariants} which 
satisfy each of the positive examples but reject each of the negative examples. 
The user chooses the appropriate invariant
and selects the sequence of corrective actions. Once all the invariants and 
their corrective actions have been generated, they are fed into the \emph{\textbf{automatic policy analyzer}}, 
which will check whether the policy has the desired effect. If the policy achieves the desired behavior, then 
it is deployed. 

\noindent\textbf{Automatic Invariant Synthesis.} The invariant synthesizer takes as input a set of positive examples, 
a set of negative examples, and a set of predicates to use in the invariant, then it tries to generate a set of candidate 
invariants that satisfy the each of the positive examples but violate each of the negative examples. The automatic invariant 
generation can be viewed as a restricted instance of the \emph{language learning from the informant problem} \cite{de2010grammatical}. 
For this, we modified an existing tool called SYSLITE \cite{arif2020syslite} to only generate invariants compliant with our policy language, by removing the capability to use the past operators not utilized in our policy language.
We use SYSLITE for invariant synthesis for its speed and scalability.

\noindent\textbf{Policy Analyses.}
The policy analyzer takes as input a policy 
and 
then helps the user check whether the policy has 
the intended behavior by performing the following two classes of analyses: 
(1) checking invariants; (2) checking the corrective 
actions. 

For class 1 policy analyses, the policy analyzer performs the following consistency checks: 
(a) does each invariant $I_j$ at least designate one state as safe? 
(b) does each invariant $I_j$ at least designate one state as unsafe? 
(c) do all invariants together at least designate one state as safe? 
(d) do all invariants together at least designate one state as unsafe? 
We solve these consistency checks by reducing it to a model checking problem in the standard way \cite{LTLSATMC}.

For class 2 policy analysis, the policy analyzer checks to see whether  corrective actions $C_j$ of 
each invariant $I_j$ can take the system to a safe state (\ie, all invariants hold) in case $I_j$ gets violated. This requires us to solve the undecidable 
decision version of the \ISSP problem.
We reduce the problem as a model 
checking problem where we use the policy checker as the model and leave the other actions (e.g., commands 
generated by automation services, human actions) as environment controlled. The guarantee this policy analysis 
provides is that there exists a path in which the system will \emph{eventually} end up in a safe state given the policy checker takes 
the necessary corrective actions.
In case execution of $C_j$ leads to violation of another invariant $I_k$, then its set of corrective actions $C_k$ should lead the system to a safe state, for $C_j$ to be accepted. 

\noindent\textbf{Policy Analysis Assumptions.} To make policy analysis tractable, we make the following assumptions: (i) the devices 
act faithfully; (ii) the communication is reliable; (iii) the number of devices are finite; (iv) the number of variables 
are finite; (v) at most one invariant violation at each state; (vi) there is only one incoming command from automation services 
or one state update from the device at any point in time.
Assumption (v) is only required for tractable policy analysis, especially,  
to simplify the underlying model checking problem to \textbf{not} maintain extra states for storing all pending corrective actions.  
Our runtime checking allows multiple invariants to be violated. 
Assumption (vi) is sound due to the fact that the underlying IoT system will \emph{serialize} concurrent events.
We leave the ordering of concurrent events as a non-deterministic choice to allow policy analysis to identify orderings leading to unsafe policy.
\section{Implementation and Evaluation}
We now discuss \approach{}'s implementation and evaluation. 

\subsection{\approach Implementation}
\label{sec:implDesign}
We develop a \approach prototype by modifying an open source MQTT broker (Mosquitto \cite{mosquitto} coded in C) to include the functionality described in Section \ref{sec:overview}.
Mosquitto acts as the \emph{automation service interface} and is responsible for consulting the policy checker whenever a command is received and forwarding commands and status updates from/to the local device interface.
We utilize MQTT for its wide support by most automation systems such as \oh, \st and Home Assistant, with an MQTT server acting as a suitable, secure intermediary with platform-agnostic capabilities.
%
%
%
The \emph{policy checker} is integrated into Mosquitto, which parses the policies during configuration time, and evaluates them on-demand 
using the approach in Appendix \ref{appendix:policyChecking}. 
\footnote{This modified version is available at https://github.com/hammadmazhar1/MAVERICK}

We utilize OpenHAB v3 \cite{openhab} as the \emph{local device interface} in our prototype with its support of WPAN communications with IoT devices and its MQTT integration capabilities.
We configure OpenHAB to expose all connected IoT devices via MQTT to Mosquitto, which then maintains a synchronzied shadow state by monitoring all device status updates sent via the device interface. 
\subsection{Data}
We use the dataset released with \helion \cite{manandhar2020Helion} to evaluate \approach under realistic settings.
\helion utilizes \emph{statistical language modeling} to learn patterns in home automation events to build a statistical language model which is then able to generate home automation event sequences that can be considered \emph{natural} by human observers.
\helion can generate both benign (\emph{up}) event sequences representing typical IoT system operation and malfunctioning (\emph{down}) event sequences where the system is behaving unexpectedly. 
%
%

\helion event sequences consist of event \emph{tokens} of the form $\langle device, capability, value\rangle$, representing an IoT \emph{device} undertaking an action resulting in \emph{value} based upon its specific \emph{capability}.
Depending upon the capability (\eg sensor, actuator), an event can occur exclusively from the device side or can occur from either direction as a cyber/physical action.
We ensure that \helion events are replayed from the relevant direction in our testbeds.
%
\subsection{Experimental Setup}
\noindent\textbf{Deployment:}
Figure \ref{fig:approachDep} illustrates our deployment architecture.
We deploy our prototype on a Raspberry Pi 4 Model B \cite{raspberrypi} running the Linux-based Raspbian OS as a local \textit{proxy} in the IoT environment, with all devices communicating to the programmable IoT platform through it.
We use 2 automation platforms -- OpenHAB and SmartThings -- for our deployment.
OpenHAB is running on a separate Raspberry Pi with a stable version (3.0.7), while we use a Samsung SmartThings Hub (STH-ETH-250) for the SmartThings platform running SmartThings V2.
%
While SmartThings does not have built-in MQTT support, we use a community-developed SmartApp \cite{smartthingsmqtt} for MQTT capabilities.
%
%
For third-party services, we select \ifttt and integrate it with the SmartThings backend using the web interface.
%
The proxy and the automation platform's backend (for \oh) or hub (for \st) are connected to the same local network.

\begin{figure}[!ht]
	\centering
	\includegraphics[scale=0.5]{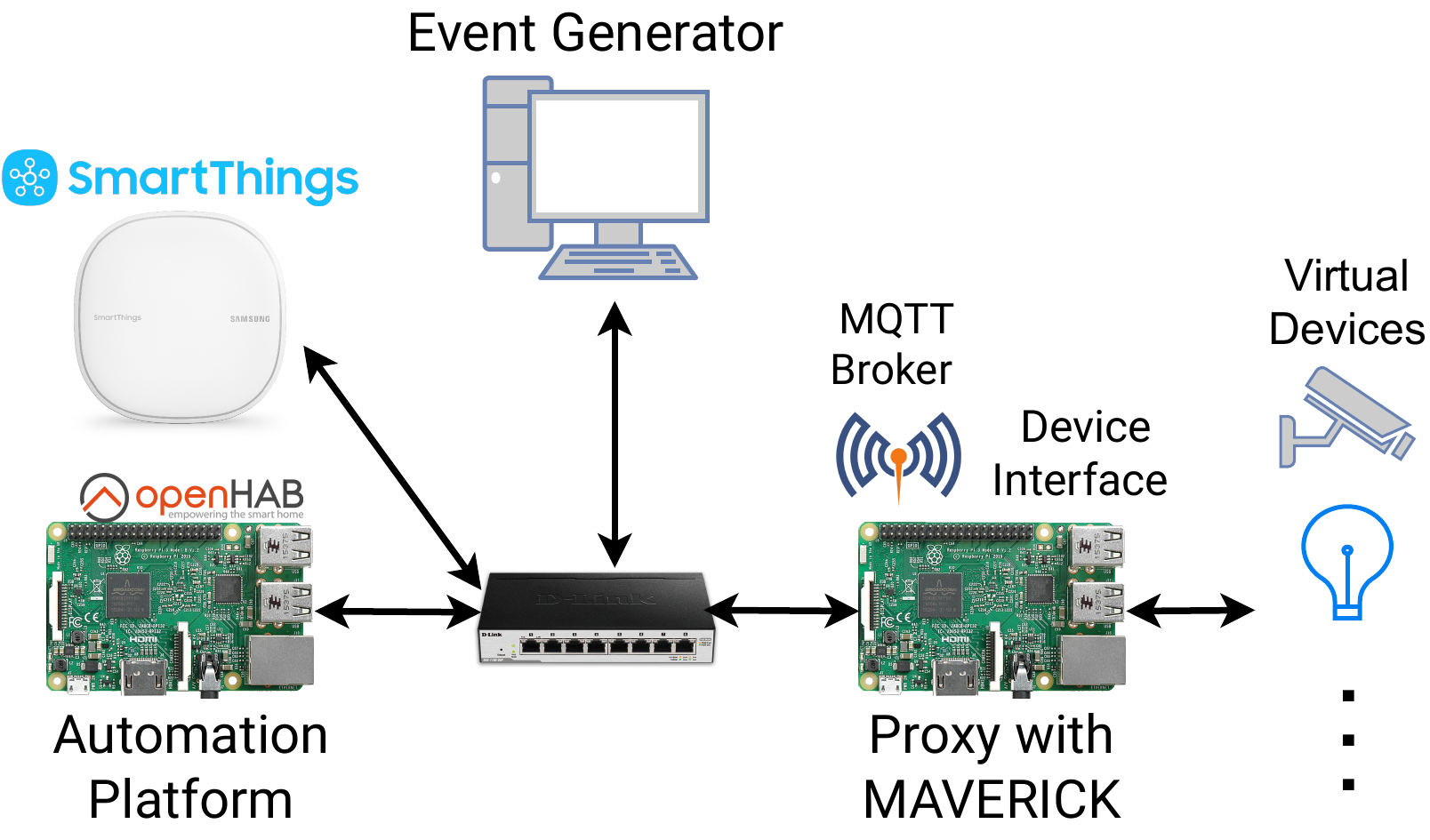}
	\caption{\approach deployment over our testbed with automation platform, event generator and virtual IoT devices.}	
	\label{fig:approachDep}
    \vspace{-0.2in}
\end{figure}

\noindent
\textbf{Testbeds:}
We prepare a virtual testbed using OpenHAB as the automation platform.
We replicate Manandhar \etal's \cite{manandhar2020Helion} virtual testbed which defined 70 home IoT devices and their capabilities, including devices such as smart locks, switches, security systems. 
%
%
We replicate device behavior virtually through the device interface on our proxy, such as change in status through commands sent by the automation platform or through physical interactions using the OpenHAB REST API.
We do not use automation apps directly in our automation platforms for most of our evaluations as \sysname's design does not require it to have prior knowledge of installed apps.
To simulate app behavior, we generate automation platform commands using their REST APIs.
This testbed allows us to virtually run event sequences generated by \helion. 
%

%

We also prepare a smaller proof-of-concept physical testbed for SmartThings to demonstrate the platform-agnostic capability of \approach.
This testbed includes only a smart lock, a smart outlet and a smart bulb.
These devices were easily connected to \approach, due to support provided by our OpenHAB-based device interface.
We ensure that devices in our physical testbed do not connect to the SmartThings hub directly so that all device commands and status updates between SmartThings and devices are routed through \approach.
%
IFTTT is integrated with this testbed to evaluate \approach's ability to intercept commands sent by third-party platforms. 
%


\begin{table*}[!t]
    \centering
    \footnotesize
\begin{tabular}{@{} r p{8cm} l l @{}} \toprule
    ID & Invariant & Corrective Action(s) & Invariant Objective \\ \midrule \midrule
    *\textsf{I1} & \textbf{If} User Away, on Vacation or Sleeping \textbf{Then} Door is Locked & (Drop Unlock Command), (Door Lock, unlock) & Physical Security \\
    *\textsf{I2} & \textbf{If} User Away, on Vacation or Sleeping \textbf{Then} Gas Stove is off & (Drop Turn on Command), (Gas Stove, off)& Physical Safety \\
    \textsf{I3} & \textbf{If} Fire Sprinkler is on \textbf{Then} Water Valve is on & (Drop Shut off Command), (Water Valve, open) & Physical Safety\\
    \textsf{I4} & \textbf{If} User Away or on Vacation \textbf{Then} Induction Cooktop is off &(Drop Turn on Command), (Induction Cooktop, off) & Physical Safety, Efficiency\\
    *\textsf{I5} & \textbf{If} User Away or on Vacation \textbf{Then} Coffee Maker is off &(Drop Turn on Command), (Coffee Maker, off) & Efficiency \\
    \textsf{I6} & \textbf{If} User Away or on Vacation Air Conditioner is off while Away or on Vacation &(Drop Turn on command), (Air Conditioner, off) & Efficiency \\
    \textsf{I7} & \textbf{If} Heater is on \textbf{Then} Air Conditioner is off & (Drop Turn on command) & Efficiency \\
    \textsf{I8} & \textbf{If} Air Conditioner is on \textbf{Then} Heater is off & (Drop Turn on command) & Efficiency \\
    \bottomrule
\end{tabular}
\vspace{-0.1in}
\caption{Policy used in evaluation. Each invariant is derived from an issue identified by the researcher in Manadhar \etal \cite{manandhar2020Helion}. Invariants marked with * are used as ground truth to measure effectiveness of policy synthesis. Corrective actions are defined by authors when configuring \approach for policy enforcement.}
\label{tab:invariants}
\vspace{-0.25in}
\end{table*}
\vspace*{-0.1in}
\subsection{Research Questions}
We evaluate the following research questions in our evaluation. 
%

\noindent\textbf{RQ1:} \emph{Can \approach synthesize invariants that match user expectations precisely?}
Recall that \approach synthesizes policy \textit{invariants} using positive and negative examples of event sequences.
We evaluate this question by providing \approach with two types of event sequences; (1) \emph{natural} event sequences generated by \helion, representing a user providing past examples of positive and negative event sequences and (2) hand-crafted \emph{synthetic} event sequences that only provide targeted positive and negative examples.

\noindent\textbf{RQ2:} \emph{How effective is \approach in enforcing user policies in IoT systems?}
We replay event sequences generated by \helion on our virtual testbed after synthesizing policies and configuring \approach with them.
As part of their evaluation, Manadhar \etal \cite{manandhar2020Helion} had a security expert analyze event sequences to identify safety issues and develop policies to mitigate them.
We use the same event sequences to evaluate \approach.

\noindent\textbf{RQ3:} \emph{What are the performance characteristics of \approach ?}
We measure \approach's performance by measuring its throughput rate, and also the latency it imposes as an intermediary. 
\subsection{Evaluation Results}

\noindent\textbf{RQ1:} We select a subset of traces from researcher-vetted \emph{down} traces from the \helion dataset that were identified with the same problem to represent natural negative examples.
For instance, in some traces the smart door lock was left unlocked while the user was away or had left on vacation, leading to a physical security issue and recommending invariant \textsf{I1} in Table \ref{tab:invariants} to safeguard against it.
Natural positive examples were generated by providing \helion with the same histories used for the negative examples, but with \helion set to \emph{up} mode.
In contrast, synthetic positive and negative examples were hand-crafted with only the relevant events in sequence.

We identify three sets of sequences of positive/negative examples for the invariants marked in Table \ref{tab:invariants}, and construct the same number of synthetic event sequences. 
We then provide these sets to \approach for invariant synthesis, asking it to generate 6 candidate invariants, noting differences between the generated and intended invariants. 
We observe that \emph{event sequence structure affects invariant stucture.}
Invariants generated via natural event sequences were more specific to the sequences compared to those generated via synthetic sequences.
This is likely because natural event sequences are non-uniform and therefore provide a larger sample space to the invariant synthesis algorithm to search through.
For natural negative sequences involving invariant \textsf{I1}, multiple events occurred between the event of the door lock unlocking and the user leaving the home, \approach synthesized invariants that required the door lock to be locked \emph{n} events before the user left, with \emph{n} dependent on the length of the provided sequences.
In contrast, with synthetic negative sequences where the door being unlocked and the user leaving had very little variance in distance, \approach was able to derive a more generalized invariant which only required the door to be locked \emph{when} the user left.   
In subsequent experiments, we use 1 invariant from the 6 candidate invariants generated for each set of synthetic event sequences, totaling 3 invariants marked in Table \ref{tab:invariants}. All other invariants were specified by the authors themselves.

We observe that \approach is able to synthesize better \emph{candidate invariants} when the user provides event sequences that precisely describe their positive and negative experiences while excluding other irrelevant events. We also observe that more diverse set of negative examples result in more generalized invariants. 
While more natural event sequences may be easier to provide through identifying periods in the system's execution history when the experiences occur, synthesizing policies using more variable-length event sequences is more difficult and can even lead to user confusion with the suggested policies.

\begin{table}[htbp]
    \centering
    \footnotesize
    \begin{tabular}{@{} p{1cm} p{5.5cm} l@{}} \toprule
        Scenario& Event Sequence & Targeted \\
        && Invariant\\ \midrule\midrule
        *Physical & Door Lock is left unlocked when user leaves. & \textsf{I1} \\
        Security& Unlock Door Lock while user is away& \\
        Physical & Turn on Gas Stove while user is away& \textsf{I2} \\
        Safety & Fire Sprinkler turns on while Water valve is closed&\textsf{I3} \\
        Efficiency & Coffee Maker is turned on while user is on vacation &\textsf{I5} \\
            & Turn on Heater while Air Conditioner is on & \textsf{I8}\\
        \bottomrule 
    \end{tabular}
    \vspace{-0.15in}
    \caption{Scenario-specific event sequences selected for evaluation, with the invariant they target. Scenarios marked with * were also evaluated on the physical testbed}
    \label{tab:scenario}
    \vspace{-0.2in}
\end{table}
\noindent\textbf{RQ2:} We select two types of sequences from the \helion dataset; (1) \emph{scenario-specific} sequences that are marked with an unsafe sequence of events by the dataset, and (2) \emph{longitudinal} sequences with longer event sequences generated using \helion.
We initialize the virtual testbed before replaying any event sequence to ensure consistency.
%
\subsubsection{Scenario-specific sequences}
Scenario-specific sequences target the set of invariants synthesized in Table \ref{tab:invariants}, with relevant events highlighted in Table \ref{tab:scenario}.
These invariants and their associated corrective actions are fed to our policy analyzer before configuration to ensure that they will lead the system to a safe state.
We use these sequences to evaluate the effectiveness of \approach in enforcing safety policies when the IoT system attempts to transition into an unsafe state.
We note \approach's behavior during each sequence to observe how it evalutes the IoT system state and reacts to unsafe states.
We observe that \approach \emph{is effective at maintaining the IoT system in a safe state when commands from the automation system could lead the system to an unsafe state.}
For instance, in the \emph{physical safety} scenario the automation system sends a command to turn on the gas stove while the user is away.
\approach evaluates the effect of the command on the current state to derive the future state which is then evaluated with the configured policy for satisfaction.
Since the resulting state does not satisfy invariant \textsf{I2}, \approach undertakes the associated corrective action \ie block the command to turn the gas stove, ensuring a safe state for the IoT system.
%

We  observe that \approach \emph{can rectify the IoT system after it transitions to an unsafe state.}
For instance, a physical safety scenario event sequence shuts off a water valve after which the fire sprinkler turns on upon detecting smoke.
Upon receiving the state update from the sprinkler device and updating its internal representation of system state, \approach's policy evaluation finds invariant \textsf{I5} to be violated by the new state.
\approach generates a corrective action to open the water valve, allowing water to flow to the sprinkler and  preventing an unsafe situation.
We observe the same in our physical testbed with the physical security scenario where the smart lock is left unlocked when the user leaves.
\approach detects that invariant \textsf{I1} is violated, and generates an corrective action to lock the door. 
\subsubsection{Longitudinal sequences}
Longitudinal sequences are generated by providing short histories (3 events) to \helion and then having it predict subsequent events (100 predictions per sequence).
We generate event sequences in both \emph{up} and \emph{down} flavors and replay them on our virtual testbed to evaluate \approach in an in-the-wild setting with the invariants in Table \ref{tab:invariants} configured.
As these sequences are run, we note the number of times \approach undertakes corrective actions, and inspect the recorded state of the testbed before and after the corrective action for correctness.
Table \ref{tab:longResults} presents the results. 
We note that across both sequence flavors, \approach initiated corrective actions for 10-14\% of events, depending upon the nature of the policy \approach is configured with during the system's operation as well as the events themselves. 
Inspection of these corrective actions revealed that most of them were generated due to violation of invariant \textsf{I1}, attempting to unlock the door after the user left the home.
In general, we note that across the longitudinal event sequences, \approach maintains the IoT system in a safe state as defined by the user's safety policy, allowing only commands that trigger safe state transitions. 
%
\begin{table}[htbp]
    \centering
    \begin{tabular}{@{} l r r r} \toprule
        Event Flavor & Events & Actions  & Correct  \\
        &   &triggered &Action Rate \\\midrule\midrule
        \emph{up (benign)}& 500 & 10 & 100\% \\
        \emph{down (malfunctioning)}&500& 70 & 100\% \\
        \bottomrule
    \end{tabular}
    \vspace{-0.1in}
    \caption{Results of running longitudinal sequences on the virtual testbed with \approach configured.}
    \label{tab:longResults}
    \vspace{-0.2in}
\end{table}

\begin{figure*}[htbp]
    \begin{subfigure}[t]{0.25\textwidth}
        \centering
        \includegraphics[width=\textwidth]{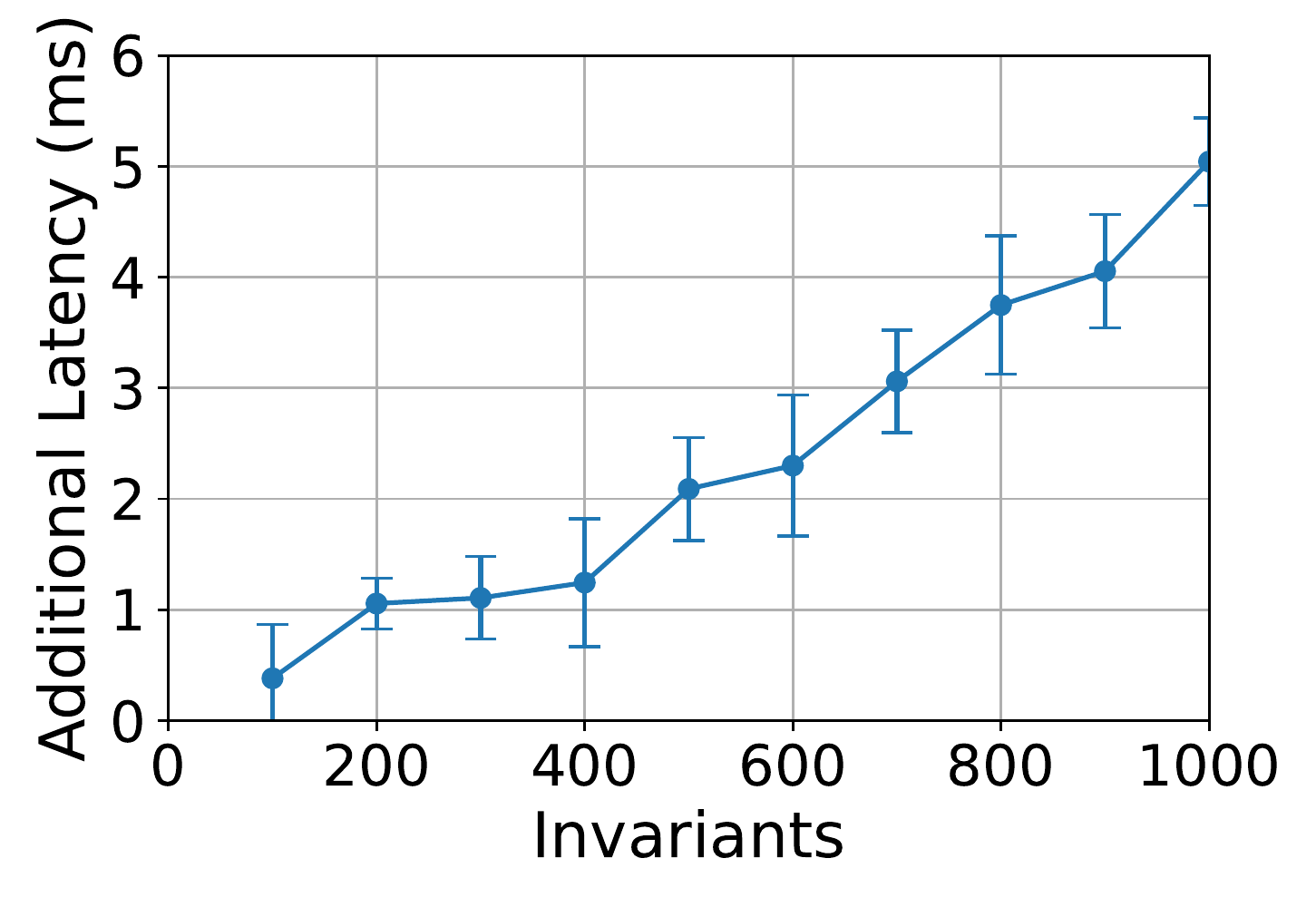}
        \vspace{-0.25in}
        \caption{Additional one-way latency}
        \label{fig:latency}
    \end{subfigure}
\begin{subfigure}[t]{0.25\textwidth}
    \centering
    \includegraphics[width=\textwidth]{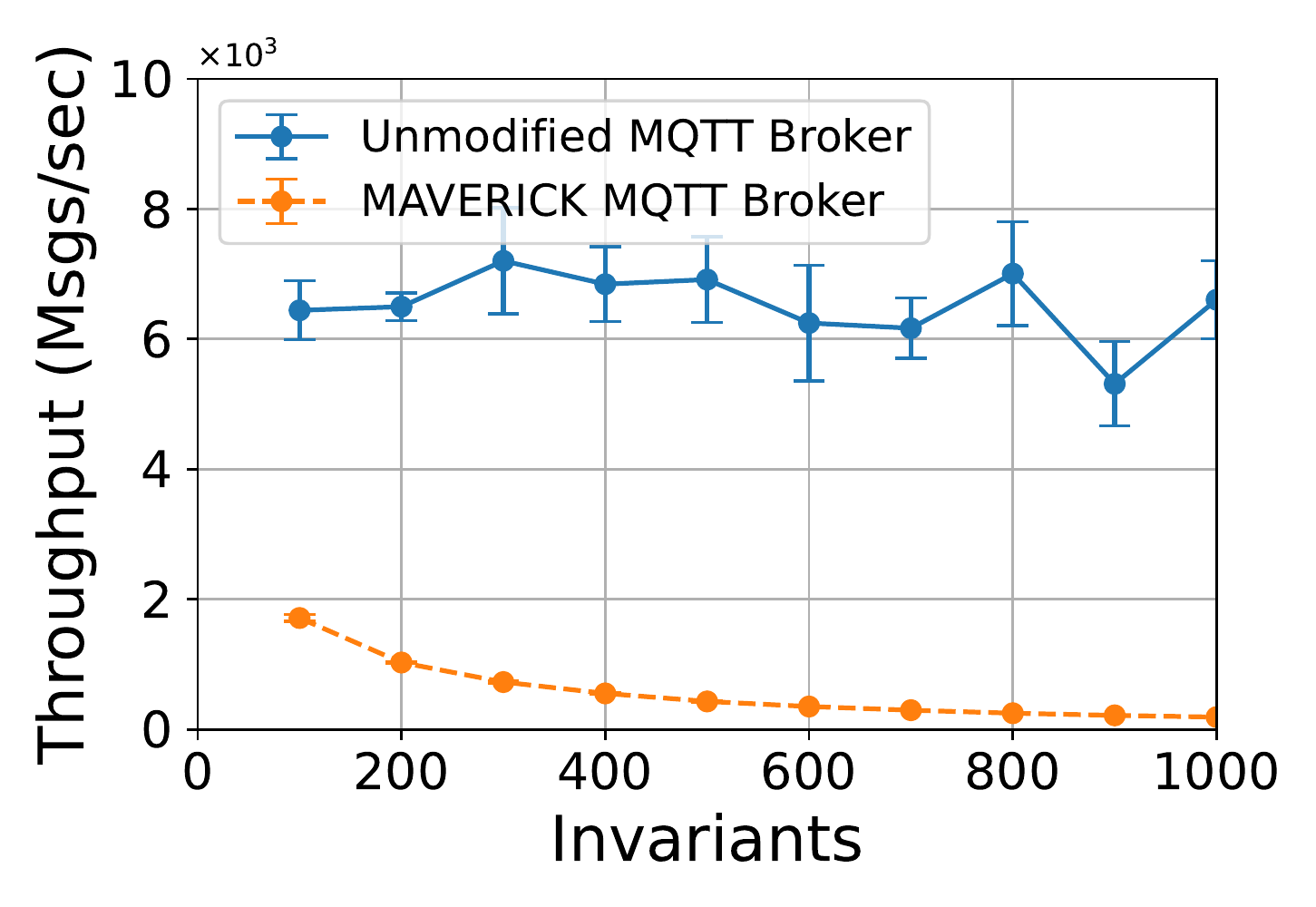}
    \vspace{-0.25in}
    \caption{Message Throughput}
    \label{fig:throughput}
\end{subfigure}
\begin{subfigure}[t]{0.25\textwidth}
    \centering
    \includegraphics[width=\textwidth]{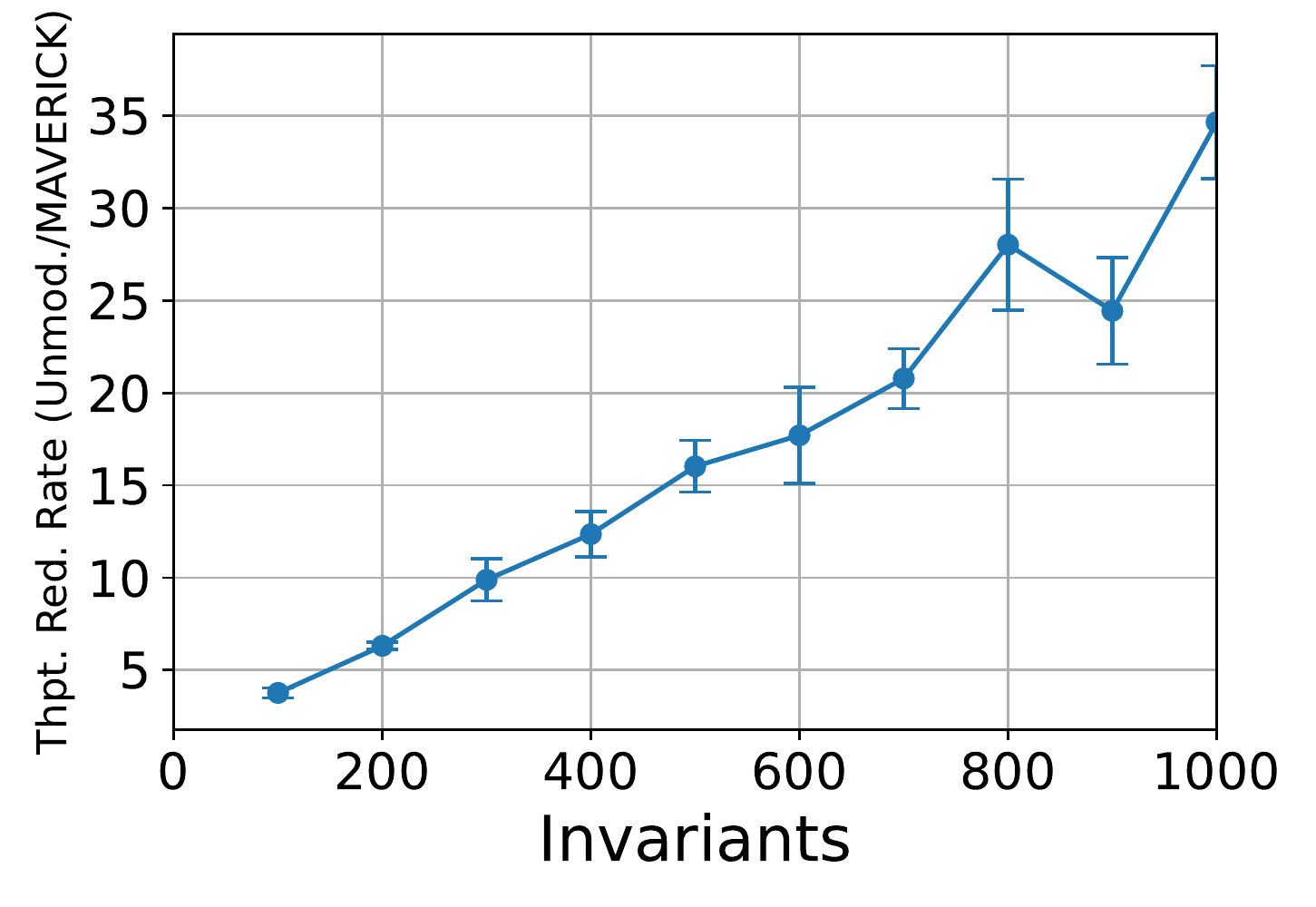}
    \vspace{-0.25in}
    \caption{Throughput Reduction Rate}
    \label{fig:thptReduction}
\end{subfigure}
\vspace{-0.25in}
\caption{Throughput and latency characteristics of \approach.}
\label{Performance}
\vspace{-0.3in}
\end{figure*}

\noindent\textbf{RQ3:} We measure performance overhead for \approach in terms of \textit{added one-way latency} and \textit{message throughput}.
We isolate \approach' and send MQTT messages through it under different sets of invariants installed.
Invariants for this evaluation are written as the simplest form; checking if a particular value (as indicated by a MQTT topic) has been set.
Therefore each invariant has an associated MQTT topic on which it applies.
We use such invariants to limit the compounding factors introduced by more complex invariants in our measurements.
 We use the same topics to send messages over an unmodified version of the broker as baseline.

Figure \ref{fig:latency} shows the average one way latency incurred by \approach.
We note that \approach{}'s latency increases with the number of invariants. 
At 1000 invariants, \approach incurs 5 milliseconds of extra latency, and the overall trend indicates that latency will increase linearly with invariant count.
Figure \ref{fig:throughput} shows the average throughput of both \approach and an unmodified MQTT broker, measured as messages per second processed by each.
In general, we observe a decrease in \approach 's throughput as the number of invariants are increased.
\approach also achieves much lower throughput as compared to the unmodified broker.
Figure \ref{fig:thptReduction} illustrates this by deriving the \textit{throughput reduction rate} caused by \approach as invariants are increased, calculated as the ratio of the unmodified MQTT broker's bandwidth over \approach.
With 100 invariants, \approach incurs ~5x throughput reduction, which increases to ~42x at 1000 invariants.

Across our measurements, we see that \approach incurs minimal \textit{latency} overhead even at relatively high invariant counts.
However, it does incur significant reduction in the message bandwidth, which may be of concern in deployments involving devices at an order of magnitude.
In such situations, it may be prudent to deploy \approach in a more localized fashion with subsets of local invariants divided across multiple \approach instances instead of a globalized deployment of a central \approach instance.
This would reduce the number of invariants a given \approach instance has to monitor at any given time.
We also note that this is just a proof-of-concept implementation of \approach and as such would likely benefit from more performance based optimizations.
One such optimization would be to pinpoint the subset of configured invariants affected by each device state variable.
Upon an update on said variable, only the identified invariants are evaluated for satisfaction, thereby further reducing the number of invariants evaluated at any time.

\vspace*{-0.15in}
\section{Discussion}\label{appendix:discussion}
%
\noindent\textbf{Invariant Expressiveness:}
\approach invariants are expressed in a fragment of 
first-order temporal logic (FOTL).  
%
While it can express \textit{temporal invariants} that refer to past states, it cannot express temporal invariants with explicit time.
As an example, \approach cannot currently capture invariants of the form: \textit{if the user has been away for at least \underline{an hour}, all security cameras must be turned on}.
\approach can be easily augmented to express and evaluate invariants of this form 
by using 
Metric-time Temporal Logic \cite{basin2015monitoring, basin2008runtime,yahyazadeh2020patriot,ozmen2022discovering} instead of FOTL.
%
This increase in expressive power, however, comes with a price as it makes both invariant synthesis and 
policy analysis intractable. 
Thus, we opt for a 
policy language that is amenable to automated invariant synthesis and policy analysis. 

\noindent\textbf{Usability:}
In order for \approach to be effective, 
it has to be usable by a typical user in IoT environments.
Deploying \approach involves 
the following two steps: (1) installing the \approach trusted intermediary in the IoT environment, and (2) configuring invariants and corrective actions.
Step (1) further involves setting up the proxy, and configuring devices and the automation system to connect to it. 
This is more involved than plug-and-play. 
%
Future  \approach versions 
will focus on making installation easier. 
%

Configuring \approach also requires writing 
invariants and corrective actions in a given policy 
language (Step 2).
%
However, an expressive policy language may not necessarily be usable for users unfamiliar with its semantics, leading to ineffective policies.
%
We address this problem by utilizing automated invariant synthesis and policy analysis techniques to assist users in developing policies.
We note that prior work does not provide such approaches, and our work is a first step in attempting to make a usable policy enforcement approach.
We also note that improving usability for such systems itself is a big challenge due to the fact that no prior knowledge of user requirements can be assumed at the time of policy specification, as well as the myraid possible system configurations possible.
%
Improved means of policy generation from automatically generated event sequences and development of User 
Interfaces which are able to guide users through the policy synthesis process would be required to improve \approach usability. 
We primarily focus on developing \approach{}'s core technology  and 
leave the evaluation of its usability 
as future work to limit scope, as developing a usable policy specification interface requires a full process of user interface development with user studies and feedback. 
\vspace*{-0.2in}
\section{Related Work}\label{sec:related}


There is a substantial amount of prior work in IoT security that focuses on device security \cite{ronen2016extended,ho2016smart,notra2014experimental,ur2013current, antonakakis2017understanding,edwards2016hajime,kumar2019iothome}, communication and authentication protocols \cite{fouladi2013honey,zigbeeFlawArticle2015,ronen2017iot,yu2015handling,gong2017piano,zhang2017proximity, barrera2018standardizing}, and IoT platforms as well as automation apps \cite{fernandes2016smart, chi2020crossapp, rahmati2018tyche,he2018rethinking, tian2017smartauth,ding2018safety,contexiot17,nguyen2018iotsan,alrawi2019sok,wang2018fear,celik2018soteria,iotfuzzer,iotguard2019ndss,lee2017fact,autotap,celik2018saint,yahyazadeh2019expat,yahyazadeh2020patriot,redini2021diane}.
Some of them aim to ensure that only policy compliant actions are allowed during a programmable IoT system execution either through static analysis 
\cite{AttackSurface, fernandes2016smart,tian2017smartauth, celik2018soteria,nguyen2018iotsan, chi2020crossapp, autotap} or through runtime monitoring \cite{iotguard2019ndss,yahyazadeh2019expat,contexiot17,yahyazadeh2020patriot,Ding2021IoTSafe,kaffle2021homeendorser}.

Static analysis based approaches that try to identify policy violations before system deployment suffer from 
the following 
limitations: (a) they suffer from imprecision due to approximate analysis, 
(b) they do not prescribe any solution to the inexperienced user on how to mitigate any identified issue, 
(c) they suffer from false positives due to lack of contextual information during analysis, and (d) modifications to the IoT environment 
(\eg, adding an app) 
warrant a new analysis.
On the contrary, \approach being a runtime approach can avoid these limitations. 
%

Prior runtime approaches aim to dynamically enforce policies, but they have multiple shortcomings: (a) They are specialized for a particular IoT platform (\ie, \st or \oh) and cannot readily be ported to other heterogeneous platforms. 
(b) They are deployed in the platform's backend (to be precise, as either \textit{instrumented apps} or \textit{instrumented APIs}).
%
Thus, third-party cloud services (like IFTTT, Flow) that use the external web API to directly communicate with IoT devices can potentially bypass these runtime approaches when these instrumented automation apps are never executed; leaving IoT devices at risk.
%
%
(c) They all expect the IoT system to never end up in an unsafe state by any environmental interference. If the system transitions to an unsafe state ever, these solutions can no longer operate. 
(d) Some of the proposed policy languages are mostly rudimentary, often platform-specific, and barely useful to express rich policies, such as temporal invariants with custom predicates.
\approach addresses all these shortcomings.  

There has been work \cite{liu2019remediot} to provide remedial actions when user expectations are violated.
Unlike \approach, they involve analysis of applications to determine when conflicts may occur, and then suggest alternative actions/rules to the user to remove the conflict.
%
%
%
%

Prior work has developed a run-time mediation of messages sent by IoT devices and automation systems to limit  
the data sent by IoT devices and thus prevent private information leakage \cite{chi2021pfirewall}.
This work has employed similar methods as ours to mediate platform-device communication, by inserting a trusted intermediary that decides what is allowed to pass between the IoT device and platform.
However, they seek to address the \textit{privacy} problem in the IoT ecosystem which is orthogonal to our efforts.
Their approach also requires analysis of apps/rules on the automation platform to ensure that 
restrictions on data release do not impact app operations.  \approach, in constrast, is completely app-agnostic. 


There have been also efforts on forensics analysis \cite{wang2018fear,wilson2017trust} and risk-based analysis \cite{ding2018safety,rahmati2018tyche,agmon2019risk}, verification of IoT device events \cite{ozmen2023evasion}, 
designing access control \cite{lee2017fact,he2018rethinking,goutam2019hestia}, repairing or synthesizing automation apps \cite{autotap}, addressing information leakage 
\cite{fernandes2016flowfence,celik2018saint,sha2018iiot}, machine learning-based anomaly detection \cite{miettinen2017iot,bezawada2018behavioral,sikder2019aegis,acar2020peek,oconnor2019homesnitch, zhang2018homonit,nguyen2019diot,pa2015iotpot,hamza2019detecting,fu2021hawatcher} and fuzzing smartphone companion apps for IoT \cite{iotfuzzer,redini2021diane}. 
These efforts are orthogonal to \approach. 

\vspace*{-0.1in}

\section{Conclusion}
%
We presented \approach, which  
ensures that deployed IoT devices 
maintain some user-defined state invariants in an app- and platform-agnostic way 
while taking physical interactions into consideration. 
%
We implemented \approach{} using a modified MQTT broker and OpenHAB and evaluated it on testbeds which showed that it can be easily deployed in IoT environments to maintain the desired state invariants in real-time.
%
Our performance evaluation also showed that while \approach may induce a lower message throughput than standard MQTT, it adds minimal latency to individual messages even when configured with a large number of invariants.
%

\begin{acks}
	The authors would like to thank Dr. Fareed Arif for his assistance in modification of SYSLITE\cite{arif2020syslite}.
	The authors would also like to thank the anonymous reviewers for their valuable comments and suggestions and the shepherd for their assistance. This material is based upon work supported by the \grantsponsor{NSFCNS2006556}{National Science Foundation}{https://www.nsf.gov/awardsearch/showAward?AWD_ID=2006556} under Grant Numbers: \grantnum{NSFCNS2006556}{CNS 2006556} and \grantnum{NSFCNS2006556}{CNS 2007512}
\end{acks}

	\bibliographystyle{ACM-Reference-Format}
	\bibliography{ref}


\begin{thebibliography}{75}


\ifx \showCODEN    \undefined \def \showCODEN     #1{\unskip}     \fi
\ifx \showDOI      \undefined \def \showDOI       #1{#1}\fi
\ifx \showISBNx    \undefined \def \showISBNx     #1{\unskip}     \fi
\ifx \showISBNxiii \undefined \def \showISBNxiii  #1{\unskip}     \fi
\ifx \showISSN     \undefined \def \showISSN      #1{\unskip}     \fi
\ifx \showLCCN     \undefined \def \showLCCN      #1{\unskip}     \fi
\ifx \shownote     \undefined \def \shownote      #1{#1}          \fi
\ifx \showarticletitle \undefined \def \showarticletitle #1{#1}   \fi
\ifx \showURL      \undefined \def \showURL       {\relax}        \fi
\providecommand\bibfield[2]{#2}
\providecommand\bibinfo[2]{#2}
\providecommand\natexlab[1]{#1}
\providecommand\showeprint[2][]{arXiv:#2}

\bibitem[Acar et~al\mbox{.}(2020)]%
        {acar2020peek}
\bibfield{author}{\bibinfo{person}{Abbas Acar}, \bibinfo{person}{Hossein
  Fereidooni}, \bibinfo{person}{Tigist Abera}, \bibinfo{person}{Amit~Kumar
  Sikder}, \bibinfo{person}{Markus Miettinen}, \bibinfo{person}{Hidayet Aksu},
  \bibinfo{person}{Mauro Conti}, \bibinfo{person}{Ahmad-Reza Sadeghi}, {and}
  \bibinfo{person}{Selcuk Uluagac}.} \bibinfo{year}{2020}\natexlab{}.
\newblock \showarticletitle{Peek-a-boo: I see your smart home activities, even
  encrypted!}. In \bibinfo{booktitle}{\emph{ACM WiSec '20}}.
\newblock


\bibitem[Agmon et~al\mbox{.}(2019)]%
        {agmon2019risk}
\bibfield{author}{\bibinfo{person}{Noga Agmon}, \bibinfo{person}{Asaf Shabtai},
  {and} \bibinfo{person}{Rami Puzis}.} \bibinfo{year}{2019}\natexlab{}.
\newblock \showarticletitle{{Deployment optimization of IoT devices through
  attack graph analysis}}. In \bibinfo{booktitle}{\emph{ACM WiSec '19}}.
\newblock


\bibitem[Alrawi et~al\mbox{.}(2019)]%
        {alrawi2019sok}
\bibfield{author}{\bibinfo{person}{Omar Alrawi}, \bibinfo{person}{Chaz Lever},
  \bibinfo{person}{Manos Antonakakis}, {and} \bibinfo{person}{Fabian Monrose}.}
  \bibinfo{year}{2019}\natexlab{}.
\newblock \showarticletitle{{SoK: Security evaluation of home-based iot
  deployments}}. In \bibinfo{booktitle}{\emph{2019 IEEE Symposium on Security
  and Privacy (S\&P)}}. IEEE.
\newblock


\bibitem[Arif et~al\mbox{.}(2020)]%
        {arif2020syslite}
\bibfield{author}{\bibinfo{person}{M.~Fareed Arif}, \bibinfo{person}{Daniel
  Larraz}, \bibinfo{person}{Mitziu Echeverria}, \bibinfo{person}{Andrew
  Reynolds}, \bibinfo{person}{Omar Chowdhury}, {and} \bibinfo{person}{Cesare
  Tinelli}.} \bibinfo{year}{2020}\natexlab{}.
\newblock \showarticletitle{SYSLITE: Syntax-Guided Synthesis of PLTL Formulas
  from Finite Traces}. In \bibinfo{booktitle}{\emph{2020 Formal Methods in
  Computer Aided Design (FMCAD)}}.
\newblock


\bibitem[Barrera et~al\mbox{.}(2018)]%
        {barrera2018standardizing}
\bibfield{author}{\bibinfo{person}{David Barrera}, \bibinfo{person}{Ian
  Molloy}, {and} \bibinfo{person}{Heqing Huang}.}
  \bibinfo{year}{2018}\natexlab{}.
\newblock \showarticletitle{Standardizing IoT network security policy
  enforcement}. In \bibinfo{booktitle}{\emph{Workshop on decentralized IoT
  security and standards (DISS)}}.
\newblock


\bibitem[Basin et~al\mbox{.}(2008)]%
        {basin2008runtime}
\bibfield{author}{\bibinfo{person}{David Basin}, \bibinfo{person}{Felix
  Klaedtke}, \bibinfo{person}{Samuel M{\"u}ller}, {and} \bibinfo{person}{Birgit
  Pfitzmann}.} \bibinfo{year}{2008}\natexlab{}.
\newblock \showarticletitle{Runtime monitoring of metric first-order temporal
  properties}. In \bibinfo{booktitle}{\emph{IARCS Annual Conference on
  Foundations of Software Technology and Theoretical Computer Science}}.
\newblock


\bibitem[Basin et~al\mbox{.}(2015)]%
        {basin2015monitoring}
\bibfield{author}{\bibinfo{person}{David Basin}, \bibinfo{person}{Felix
  Klaedtke}, \bibinfo{person}{Samuel M{\"u}ller}, {and} \bibinfo{person}{Eugen
  Z{\u{a}}linescu}.} \bibinfo{year}{2015}\natexlab{}.
\newblock \showarticletitle{Monitoring metric first-order temporal properties}.
\newblock \bibinfo{journal}{\emph{Journal of the ACM (JACM)}}
  (\bibinfo{year}{2015}).
\newblock


\bibitem[Basin et~al\mbox{.}(2017)]%
        {monpoly}
\bibfield{author}{\bibinfo{person}{David~A Basin}, \bibinfo{person}{Felix
  Klaedtke}, {and} \bibinfo{person}{Eugen Zalinescu}.}
  \bibinfo{year}{2017}\natexlab{}.
\newblock \showarticletitle{The {MonPoly} Monitoring Tool.}
\newblock \bibinfo{journal}{\emph{RV-CuBES}} (\bibinfo{year}{2017}).
\newblock


\bibitem[Bezawada et~al\mbox{.}(2018)]%
        {bezawada2018behavioral}
\bibfield{author}{\bibinfo{person}{Bruhadeshwar Bezawada},
  \bibinfo{person}{Maalvika Bachani}, \bibinfo{person}{Jordan Peterson},
  \bibinfo{person}{Hossein Shirazi}, \bibinfo{person}{Indrakshi Ray}, {and}
  \bibinfo{person}{Indrajit Ray}.} \bibinfo{year}{2018}\natexlab{}.
\newblock \showarticletitle{Behavioral fingerprinting of iot devices}. In
  \bibinfo{booktitle}{\emph{Proceedings of the 2018 workshop on attacks and
  solutions in hardware security}}.
\newblock


\bibitem[Celik et~al\mbox{.}(2018a)]%
        {celik2018saint}
\bibfield{author}{\bibinfo{person}{Z.~Berkay Celik}, \bibinfo{person}{Leonardo
  Babun}, \bibinfo{person}{Amit~Kumar Sikder}, \bibinfo{person}{Hidayet Aksu},
  \bibinfo{person}{Gang Tan}, \bibinfo{person}{Patrick McDaniel}, {and}
  \bibinfo{person}{A.~Selcuk Uluagac}.} \bibinfo{year}{2018}\natexlab{a}.
\newblock \showarticletitle{Sensitive Information Tracking in Commodity IoT}.
  In \bibinfo{booktitle}{\emph{27th {USENIX} Security Symposium ({USENIX}
  Security 18)}}.
\newblock


\bibitem[Celik et~al\mbox{.}(2018b)]%
        {celik2018soteria}
\bibfield{author}{\bibinfo{person}{Z~Berkay Celik}, \bibinfo{person}{Patrick
  McDaniel}, {and} \bibinfo{person}{Gang Tan}.}
  \bibinfo{year}{2018}\natexlab{b}.
\newblock \showarticletitle{Soteria: Automated {IoT} Safety and Security
  Analysis}. In \bibinfo{booktitle}{\emph{2018 USENIX Annual Technical
  Conference (USENIX ATC 18)}}.
\newblock


\bibitem[Celik et~al\mbox{.}(2019)]%
        {iotguard2019ndss}
\bibfield{author}{\bibinfo{person}{Z~Berkay Celik}, \bibinfo{person}{Gang Tan},
  {and} \bibinfo{person}{Patrick~D McDaniel}.} \bibinfo{year}{2019}\natexlab{}.
\newblock \showarticletitle{IoTGuard: Dynamic Enforcement of Security and
  Safety Policy in Commodity IoT.}. In \bibinfo{booktitle}{\emph{NDSS}}.
\newblock


\bibitem[Chen et~al\mbox{.}(2018)]%
        {iotfuzzer}
\bibfield{author}{\bibinfo{person}{Jiongyi Chen}, \bibinfo{person}{Wenrui
  Diao}, \bibinfo{person}{Qingchuan Zhao}, \bibinfo{person}{Chaoshun Zuo},
  \bibinfo{person}{Zhiqiang Lin}, \bibinfo{person}{XiaoFeng Wang},
  \bibinfo{person}{Wing~Cheong Lau}, \bibinfo{person}{Menghan Sun},
  \bibinfo{person}{Ronghai Yang}, {and} \bibinfo{person}{Kehuan Zhang}.}
  \bibinfo{year}{2018}\natexlab{}.
\newblock \showarticletitle{IoTFuzzer: Discovering Memory Corruptions in IoT
  Through App-based Fuzzing.}. In \bibinfo{booktitle}{\emph{NDSS}}.
\newblock


\bibitem[Chi et~al\mbox{.}(2021)]%
        {chi2021pfirewall}
\bibfield{author}{\bibinfo{person}{Haotian Chi}, \bibinfo{person}{Qiang Zeng},
  \bibinfo{person}{Xiaojiang Du}, {and} \bibinfo{person}{Lannan Luo}.}
  \bibinfo{year}{2021}\natexlab{}.
\newblock \showarticletitle{PFirewall: Semantics-Aware Customizable Data Flow
  Control for Smart Home Privacy Protection}.
\newblock \bibinfo{journal}{\emph{Network and Distributed System Security
  Symposium}}.
\newblock


\bibitem[Chi et~al\mbox{.}(2020)]%
        {chi2020crossapp}
\bibfield{author}{\bibinfo{person}{Haotian Chi}, \bibinfo{person}{Qiang Zeng},
  \bibinfo{person}{Xiaojiang Du}, {and} \bibinfo{person}{Jiaping Yu}.}
  \bibinfo{year}{2020}\natexlab{}.
\newblock \showarticletitle{Cross-app interference threats in smart homes:
  Categorization, detection and handling}. In \bibinfo{booktitle}{\emph{2020
  50th Annual IEEE/IFIP International Conference on Dependable Systems and
  Networks (DSN)}}. IEEE.
\newblock


\bibitem[De~la Higuera(2010)]%
        {de2010grammatical}
\bibfield{author}{\bibinfo{person}{Colin De~la Higuera}.}
  \bibinfo{year}{2010}\natexlab{}.
\newblock \bibinfo{booktitle}{\emph{Grammatical inference: learning automata
  and grammars}}.
\newblock \bibinfo{publisher}{Cambridge University Press}.
\newblock


\bibitem[Ding and Hu(2018)]%
        {ding2018safety}
\bibfield{author}{\bibinfo{person}{Wenbo Ding} {and} \bibinfo{person}{Hongxin
  Hu}.} \bibinfo{year}{2018}\natexlab{}.
\newblock \showarticletitle{On the safety of iot device physical interaction
  control}. In \bibinfo{booktitle}{\emph{Proceedings of the 2018 ACM SIGSAC
  Conference on Computer and Communications Security}}.
\newblock


\bibitem[Ding et~al\mbox{.}(2021)]%
        {Ding2021IoTSafe}
\bibfield{author}{\bibinfo{person}{Wenbo Ding}, \bibinfo{person}{Hongxin Hu},
  {and} \bibinfo{person}{Long Cheng}.} \bibinfo{year}{2021}\natexlab{}.
\newblock \showarticletitle{IoTSafe: Enforcing Safety and Security Policy with
  Real IoT Physical Interaction Discovery}. In
  \bibinfo{booktitle}{\emph{NDSS}}.
\newblock


\bibitem[Echeverria et~al\mbox{.}(2021)]%
        {echeverriaphoenix}
\bibfield{author}{\bibinfo{person}{Mitziu Echeverria}, \bibinfo{person}{Zeeshan
  Ahmed}, \bibinfo{person}{Bincheng Wang}, \bibinfo{person}{M~Fareed Arif},
  \bibinfo{person}{Syed~Rafiul Hussain}, {and} \bibinfo{person}{Omar
  Chowdhury}.} \bibinfo{year}{2021}\natexlab{}.
\newblock \showarticletitle{PHOENIX: Device-Centric Cellular Network Protocol
  Monitoring using Runtime Verification}. In
  \bibinfo{booktitle}{\emph{Proceedings of the 2021 Network and Distributed
  Systems Symposium}}.
\newblock


\bibitem[Edwards and Profetis(2016)]%
        {edwards2016hajime}
\bibfield{author}{\bibinfo{person}{Sam Edwards} {and} \bibinfo{person}{Ioannis
  Profetis}.} \bibinfo{year}{2016}\natexlab{}.
\newblock \showarticletitle{Hajime: Analysis of a decentralized internet worm
  for IoT devices}.
\newblock \bibinfo{journal}{\emph{Rapidity Networks}} (\bibinfo{year}{2016}).
\newblock


\bibitem[Ergen(2004)]%
        {zigbee}
\bibfield{author}{\bibinfo{person}{Sinem~Coleri Ergen}.}
  \bibinfo{year}{2004}\natexlab{}.
\newblock \showarticletitle{ZigBee/IEEE 802.15. 4 Summary}.
\newblock \bibinfo{journal}{\emph{UC Berkeley, September}}
  \bibinfo{volume}{10}, \bibinfo{number}{17} (\bibinfo{year}{2004}),
  \bibinfo{pages}{11}.
\newblock


\bibitem[Fernandes et~al\mbox{.}(2016a)]%
        {fernandes2016smart}
\bibfield{author}{\bibinfo{person}{E. Fernandes}, \bibinfo{person}{J. Jung},
  {and} \bibinfo{person}{A. Prakash}.} \bibinfo{year}{2016}\natexlab{a}.
\newblock \showarticletitle{Security Analysis of Emerging Smart Home
  Applications}. In \bibinfo{booktitle}{\emph{2016 IEEE Symposium on Security
  and Privacy (S\&P)}}. \bibinfo{publisher}{IEEE}.
\newblock


\bibitem[Fernandes et~al\mbox{.}(2016b)]%
        {fernandes2016flowfence}
\bibfield{author}{\bibinfo{person}{Earlence Fernandes}, \bibinfo{person}{Justin
  Paupore}, \bibinfo{person}{Amir Rahmati}, \bibinfo{person}{Daniel Simionato},
  \bibinfo{person}{Mauro Conti}, {and} \bibinfo{person}{Atul Prakash}.}
  \bibinfo{year}{2016}\natexlab{b}.
\newblock \showarticletitle{{FlowFence}: Practical Data Protection for Emerging
  IoT Application Frameworks}. In \bibinfo{booktitle}{\emph{USENIX Security
  '16}}.
\newblock


\bibitem[{FM Global}(2016)]%
        {ssprinkler}
\bibfield{author}{\bibinfo{person}{{FM Global}}.}
  \bibinfo{year}{2016}\natexlab{}.
\newblock \bibinfo{booktitle}{\emph{{UNCOMMON SENSORS : Researchers pioneer
  smart sprinklers, with greater detection accuracy and speed of response}}}.
\newblock


\bibitem[Fouladi and Ghanoun(2013)]%
        {fouladi2013honey}
\bibfield{author}{\bibinfo{person}{Behrang Fouladi} {and}
  \bibinfo{person}{Sahand Ghanoun}.} \bibinfo{year}{2013}\natexlab{}.
\newblock \showarticletitle{Honey, I'm Home!!, Hacking ZWave Home Automation
  Systems}.
\newblock \bibinfo{journal}{\emph{Black Hat USA}} (\bibinfo{year}{2013}).
\newblock


\bibitem[Fu et~al\mbox{.}(2021)]%
        {fu2021hawatcher}
\bibfield{author}{\bibinfo{person}{Chenglong Fu}, \bibinfo{person}{Qiang Zeng},
  {and} \bibinfo{person}{Xiaojiang Du}.} \bibinfo{year}{2021}\natexlab{}.
\newblock \showarticletitle{HAWatcher: Semantics-aware anomaly detection for
  appified smart homes}. In \bibinfo{booktitle}{\emph{30th $\{$USENIX$\}$
  Security Symposium ($\{$USENIX$\}$ Security 21)}}.
\newblock


\bibitem[Gong et~al\mbox{.}(2017)]%
        {gong2017piano}
\bibfield{author}{\bibinfo{person}{Neil~Zhenqiang Gong}, \bibinfo{person}{Altay
  Ozen}, \bibinfo{person}{Yu Wu}, \bibinfo{person}{Xiaoyu Cao},
  \bibinfo{person}{Richard Shin}, \bibinfo{person}{Dawn Song},
  \bibinfo{person}{Hongxia Jin}, {and} \bibinfo{person}{Xuan Bao}.}
  \bibinfo{year}{2017}\natexlab{}.
\newblock \showarticletitle{PIANO: Proximity-based user authentication on
  voice-powered internet-of-things devices}. In \bibinfo{booktitle}{\emph{2017
  IEEE 37th International Conference on Distributed Computing Systems
  (ICDCS)}}. IEEE.
\newblock


\bibitem[Goutam et~al\mbox{.}(2019)]%
        {goutam2019hestia}
\bibfield{author}{\bibinfo{person}{Sanket Goutam}, \bibinfo{person}{William
  Enck}, {and} \bibinfo{person}{Bradley Reaves}.}
  \bibinfo{year}{2019}\natexlab{}.
\newblock \showarticletitle{Hestia: simple least privilege network policies for
  smart homes}. In \bibinfo{booktitle}{\emph{Proceedings of the 12th Conference
  on Security and Privacy in Wireless and Mobile Networks}}.
  \bibinfo{pages}{215--220}.
\newblock


\bibitem[Gupta(2016)]%
        {gupta2016inside}
\bibfield{author}{\bibinfo{person}{Naresh~Kumar Gupta}.}
  \bibinfo{year}{2016}\natexlab{}.
\newblock \bibinfo{booktitle}{\emph{Inside Bluetooth low energy}}.
\newblock \bibinfo{publisher}{Artech House}.
\newblock


\bibitem[Hamza et~al\mbox{.}(2019)]%
        {hamza2019detecting}
\bibfield{author}{\bibinfo{person}{Ayyoob Hamza},
  \bibinfo{person}{Hassan~Habibi Gharakheili}, \bibinfo{person}{Theophilus~A
  Benson}, {and} \bibinfo{person}{Vijay Sivaraman}.}
  \bibinfo{year}{2019}\natexlab{}.
\newblock \showarticletitle{Detecting volumetric attacks on lot devices via
  sdn-based monitoring of mud activity}. In
  \bibinfo{booktitle}{\emph{Proceedings of the 2019 ACM Symposium on SDN
  Research}}.
\newblock


\bibitem[Harrison et~al\mbox{.}(1976)]%
        {harrison1976protection}
\bibfield{author}{\bibinfo{person}{Michael~A Harrison},
  \bibinfo{person}{Walter~L Ruzzo}, {and} \bibinfo{person}{Jeffrey~D Ullman}.}
  \bibinfo{year}{1976}\natexlab{}.
\newblock \showarticletitle{Protection in operating systems}.
\newblock \bibinfo{journal}{\emph{Commun. ACM}} (\bibinfo{year}{1976}).
\newblock


\bibitem[He et~al\mbox{.}(2018)]%
        {he2018rethinking}
\bibfield{author}{\bibinfo{person}{Weijia He}, \bibinfo{person}{Maximilian
  Golla}, \bibinfo{person}{Roshni Padhi}, \bibinfo{person}{Jordan Ofek},
  \bibinfo{person}{Markus D{\"u}rmuth}, \bibinfo{person}{Earlence Fernandes},
  {and} \bibinfo{person}{Blase Ur}.} \bibinfo{year}{2018}\natexlab{}.
\newblock \showarticletitle{Rethinking Access Control and Authentication for
  the Home Internet of Things {(IoT)}}. In \bibinfo{booktitle}{\emph{27th
  USENIX Security Symposium (USENIX Security 18)}}.
\newblock


\bibitem[Ho et~al\mbox{.}(2016)]%
        {ho2016smart}
\bibfield{author}{\bibinfo{person}{Grant Ho}, \bibinfo{person}{Derek Leung},
  \bibinfo{person}{Pratyush Mishra}, \bibinfo{person}{Ashkan Hosseini},
  \bibinfo{person}{Dawn Song}, {and} \bibinfo{person}{David Wagner}.}
  \bibinfo{year}{2016}\natexlab{}.
\newblock \showarticletitle{Smart locks: Lessons for securing commodity
  internet of things devices}. In \bibinfo{booktitle}{\emph{Proceedings of the
  11th ACM on Asia conference on computer and communications security}}.
\newblock


\bibitem[howpublishedkakis et~al\mbox{.}(2017)]%
        {antonakakis2017understanding}
\bibfield{author}{\bibinfo{person}{Manos howpublishedkakis},
  \bibinfo{person}{Tim April}, \bibinfo{person}{Michael Bailey},
  \bibinfo{person}{Matt Bernhard}, \bibinfo{person}{Elie Bursztein},
  \bibinfo{person}{Jaime Cochran}, \bibinfo{person}{Zakir Durumeric},
  \bibinfo{person}{J~Alex Halderman}, \bibinfo{person}{Luca Invernizzi},
  \bibinfo{person}{Michalis Kallitsis}, {et~al\mbox{.}}}
  \bibinfo{year}{2017}\natexlab{}.
\newblock \showarticletitle{{Understanding the Mirai botnet}}. In
  \bibinfo{booktitle}{\emph{26th USENIX security symposium (USENIX Security
  17)}}.
\newblock


\bibitem[{IFTTT}(2022)]%
        {wpa}
\bibfield{author}{\bibinfo{person}{{IFTTT}}.} \bibinfo{year}{2022}\natexlab{}.
\newblock \bibinfo{booktitle}{\emph{Turn water off if D-Link water sensoor
  detects water}}.
\newblock


\bibitem[IFTTT(2023)]%
        {IFTTTfiltercode}
\bibfield{author}{\bibinfo{person}{IFTTT}.} \bibinfo{year}{2023}\natexlab{}.
\newblock \bibinfo{booktitle}{\emph{Filter Code - IFTTT Help Center}}.
\newblock


\bibitem[Jia et~al\mbox{.}(2017)]%
        {contexiot17}
\bibfield{author}{\bibinfo{person}{Yunhan~Jack Jia}, \bibinfo{person}{Qi~Alfred
  Chen}, \bibinfo{person}{Shiqi Wang}, \bibinfo{person}{Amir Rahmati},
  \bibinfo{person}{Earlence Fernandes}, \bibinfo{person}{Z.~Morley Mao}, {and}
  \bibinfo{person}{Atul Prakash}.} \bibinfo{year}{2017}\natexlab{}.
\newblock \showarticletitle{{ContexIoT: Towards Providing Contextual Integrity
  to Appified IoT Platforms}}. In \bibinfo{booktitle}{\emph{21st Network and
  Distributed Security Symposium (NDSS)}}.
\newblock


\bibitem[Johnson(2021)]%
        {smartthingsmqtt}
\bibfield{author}{\bibinfo{person}{St.~John Johnson}.}
  \bibinfo{year}{2021}\natexlab{}.
\newblock \bibinfo{booktitle}{\emph{SmartThings MQTT Bridge}}.
\newblock


\bibitem[Kafle et~al\mbox{.}(2021)]%
        {kaffle2021homeendorser}
\bibfield{author}{\bibinfo{person}{Kaushal Kafle}, \bibinfo{person}{Kirti
  Jagtap}, \bibinfo{person}{Mansoor Ahmed-Rengers}, \bibinfo{person}{Trent
  Jaeger}, {and} \bibinfo{person}{Adwait Nadkarni}.}
  \bibinfo{year}{2021}\natexlab{}.
\newblock \bibinfo{title}{Towards Practical Integrity in the Smart Home with
  HomeEndorser}.
\newblock
\newblock
\urldef\tempurl%
\url{https://doi.org/10.48550/ARXIV.2109.05139}
\showDOI{\tempurl}


\bibitem[Kumar et~al\mbox{.}(2019)]%
        {kumar2019iothome}
\bibfield{author}{\bibinfo{person}{Deepak Kumar}, \bibinfo{person}{Kelly Shen},
  \bibinfo{person}{Benton Case}, \bibinfo{person}{Deepali Garg},
  \bibinfo{person}{Galina Alperovich}, \bibinfo{person}{Dmitry Kuznetsov},
  \bibinfo{person}{Rajarshi Gupta}, {and} \bibinfo{person}{Zakir Durumeric}.}
  \bibinfo{year}{2019}\natexlab{}.
\newblock \showarticletitle{All Things Considered: An Analysis of IoT Devices
  on Home Networks}. In \bibinfo{booktitle}{\emph{28th {USENIX} Security
  Symposium ({USENIX} Security 19)}}. \bibinfo{publisher}{{USENIX}
  Association}.
\newblock


\bibitem[Lee et~al\mbox{.}(2017)]%
        {lee2017fact}
\bibfield{author}{\bibinfo{person}{Sanghak Lee}, \bibinfo{person}{Jiwon Choi},
  \bibinfo{person}{Jihun Kim}, \bibinfo{person}{Beumjin Cho},
  \bibinfo{person}{Sangho Lee}, \bibinfo{person}{Hanjun Kim}, {and}
  \bibinfo{person}{Jong Kim}.} \bibinfo{year}{2017}\natexlab{}.
\newblock \showarticletitle{FACT: Functionality-centric access control system
  for IoT programming frameworks}. In \bibinfo{booktitle}{\emph{Proceedings of
  the 22nd ACM on Symposium on Access Control Models and Technologies}}. ACM.
\newblock


\bibitem[Liu et~al\mbox{.}(2019)]%
        {liu2019remediot}
\bibfield{author}{\bibinfo{person}{Renju Liu}, \bibinfo{person}{Ziqi Wang},
  \bibinfo{person}{Luis Garcia}, {and} \bibinfo{person}{Mani Srivastava}.}
  \bibinfo{year}{2019}\natexlab{}.
\newblock \showarticletitle{RemedIoT: Remedial actions for Internet-of-Things
  conflicts}. In \bibinfo{booktitle}{\emph{Proceedings of the 6th ACM
  International Conference on Systems for Energy-Efficient Buildings, Cities,
  and Transportation}}.
\newblock


\bibitem[Lomas(2015)]%
        {zigbeeFlawArticle2015}
\bibfield{author}{\bibinfo{person}{Natasha Lomas}.}
  \bibinfo{year}{2015}\natexlab{}.
\newblock \bibinfo{title}{Critical Flaw IDed In ZigBee Smart Home Devices}.
\newblock
  \bibinfo{howpublished}{\url{https://techcrunch.com/2015/08/07/critical-flaw-ided-in-zigbee-smart-home-devices/}}.
\newblock


\bibitem[Manandhar et~al\mbox{.}(2020)]%
        {manandhar2020Helion}
\bibfield{author}{\bibinfo{person}{Sunil Manandhar}, \bibinfo{person}{Kevin
  Moran}, \bibinfo{person}{Kaushal Kafle}, \bibinfo{person}{Ruhao Tang},
  \bibinfo{person}{Denys Poshyvanyk}, {and} \bibinfo{person}{Adwait Nadkarni}.}
  \bibinfo{year}{2020}\natexlab{}.
\newblock \showarticletitle{Towards a natural perspective of smart homes for
  practical security and safety analyses}. In \bibinfo{booktitle}{\emph{2020
  IEEE Symposium on Security and Privacy (S\&P)}}. IEEE.
\newblock


\bibitem[Miettinen et~al\mbox{.}(2017)]%
        {miettinen2017iot}
\bibfield{author}{\bibinfo{person}{Markus Miettinen}, \bibinfo{person}{Samuel
  Marchal}, \bibinfo{person}{Ibbad Hafeez}, \bibinfo{person}{N Asokan},
  \bibinfo{person}{Ahmad-Reza Sadeghi}, {and} \bibinfo{person}{Sasu Tarkoma}.}
  \bibinfo{year}{2017}\natexlab{}.
\newblock \showarticletitle{IoT Sentinel: Automated device-type identification
  for security enforcement in IoT}. In \bibinfo{booktitle}{\emph{2017 IEEE 37th
  International Conference on Distributed Computing Systems (ICDCS)}}. IEEE.
\newblock


\bibitem[Mosquitto(2022)]%
        {mosquitto}
\bibfield{author}{\bibinfo{person}{Mosquitto}.}
  \bibinfo{year}{2022}\natexlab{}.
\newblock \bibinfo{booktitle}{\emph{Eclipse Mosquitto}}.
\newblock


\bibitem[Nguyen et~al\mbox{.}(2018)]%
        {nguyen2018iotsan}
\bibfield{author}{\bibinfo{person}{Dang~Tu Nguyen}, \bibinfo{person}{Chengyu
  Song}, \bibinfo{person}{Zhiyun Qian}, \bibinfo{person}{Srikanth~V
  Krishnamurthy}, \bibinfo{person}{Edward~JM Colbert}, {and}
  \bibinfo{person}{Patrick McDaniel}.} \bibinfo{year}{2018}\natexlab{}.
\newblock \showarticletitle{IotSan: fortifying the safety of IoT systems}. In
  \bibinfo{booktitle}{\emph{Proceedings of the 14th International Conference on
  emerging Networking EXperiments and Technologies}}. ACM.
\newblock


\bibitem[Nguyen et~al\mbox{.}(2019)]%
        {nguyen2019diot}
\bibfield{author}{\bibinfo{person}{Thien~Duc Nguyen}, \bibinfo{person}{Samuel
  Marchal}, \bibinfo{person}{Markus Miettinen}, \bibinfo{person}{Hossein
  Fereidooni}, \bibinfo{person}{N Asokan}, {and} \bibinfo{person}{Ahmad-Reza
  Sadeghi}.} \bibinfo{year}{2019}\natexlab{}.
\newblock \showarticletitle{D{\"I}oT: A federated self-learning anomaly
  detection system for IoT}. In \bibinfo{booktitle}{\emph{2019 IEEE 39th
  International Conference on Distributed Computing Systems (ICDCS)}}. IEEE.
\newblock


\bibitem[Notra et~al\mbox{.}(2014)]%
        {notra2014experimental}
\bibfield{author}{\bibinfo{person}{Sukhvir Notra}, \bibinfo{person}{Muhammad
  Siddiqi}, \bibinfo{person}{Hassan~Habibi Gharakheili}, \bibinfo{person}{Vijay
  Sivaraman}, {and} \bibinfo{person}{Roksana Boreli}.}
  \bibinfo{year}{2014}\natexlab{}.
\newblock \showarticletitle{An experimental study of security and privacy risks
  with emerging household appliances}. In \bibinfo{booktitle}{\emph{2014 IEEE
  Conference on Communications and Network Security (CNS)}}. IEEE.
\newblock


\bibitem[OConnor et~al\mbox{.}(2019)]%
        {oconnor2019homesnitch}
\bibfield{author}{\bibinfo{person}{TJ OConnor}, \bibinfo{person}{Reham
  Mohamed}, \bibinfo{person}{Markus Miettinen}, \bibinfo{person}{William Enck},
  \bibinfo{person}{Bradley Reaves}, {and} \bibinfo{person}{Ahmad-Reza
  Sadeghi}.} \bibinfo{year}{2019}\natexlab{}.
\newblock \showarticletitle{HomeSnitch: behavior transparency and control for
  smart home IoT devices}. In \bibinfo{booktitle}{\emph{Proceedings of the 12th
  Conference on Security and Privacy in Wireless and Mobile Networks}}. ACM.
\newblock


\bibitem[openHAB(2019)]%
        {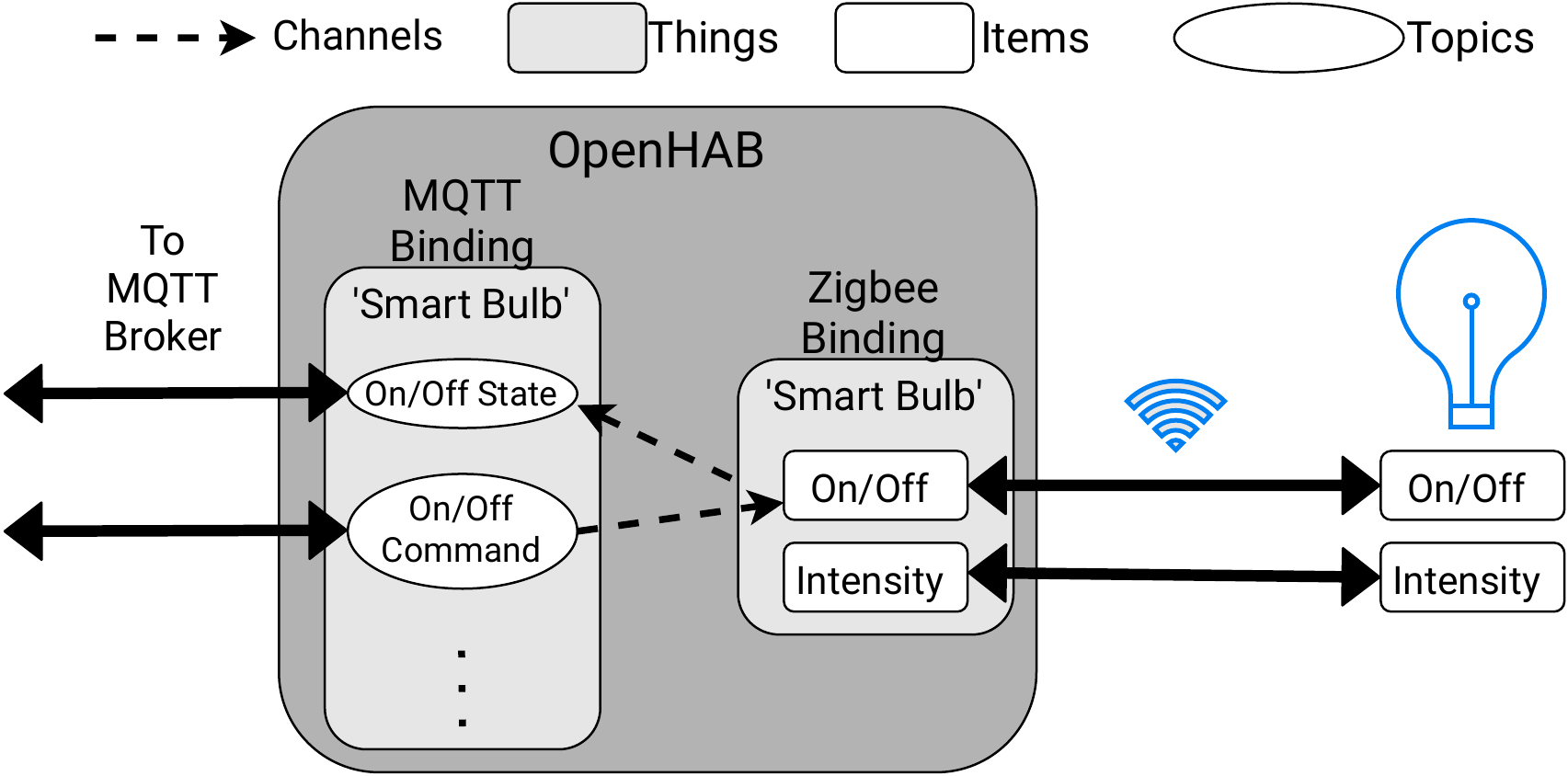}
\bibfield{author}{\bibinfo{person}{openHAB}.} \bibinfo{year}{2019}\natexlab{}.
\newblock \bibinfo{booktitle}{}.
\newblock


\bibitem[Ozmen et~al\mbox{.}(2022)]%
        {ozmen2022discovering}
\bibfield{author}{\bibinfo{person}{Muslum~Ozgur Ozmen},
  \bibinfo{person}{Xuansong Li}, \bibinfo{person}{Andrew Chu},
  \bibinfo{person}{Z~Berkay Celik}, \bibinfo{person}{Bardh Hoxha}, {and}
  \bibinfo{person}{Xiangyu Zhang}.} \bibinfo{year}{2022}\natexlab{}.
\newblock \showarticletitle{Discovering IoT Physical Channel Vulnerabilities}.
  In \bibinfo{booktitle}{\emph{Proceedings of the 2022 ACM SIGSAC Conference on
  Computer and Communications Security}}.
\newblock


\bibitem[Ozmen et~al\mbox{.}(2023)]%
        {ozmen2023evasion}
\bibfield{author}{\bibinfo{person}{Muslum~Ozgur Ozmen}, \bibinfo{person}{Ruoyu
  Song}, \bibinfo{person}{Habiba Farrukh}, {and} \bibinfo{person}{Z~Berkay
  Celik}.} \bibinfo{year}{2023}\natexlab{}.
\newblock \showarticletitle{Evasion attacks and defenses on smart home physical
  event verification}. In \bibinfo{booktitle}{\emph{Network and Distributed
  System Security (NDSS)'23}}.
\newblock


\bibitem[Pa et~al\mbox{.}(2015)]%
        {pa2015iotpot}
\bibfield{author}{\bibinfo{person}{Yin Minn~Pa Pa}, \bibinfo{person}{Shogo
  Suzuki}, \bibinfo{person}{Katsunari Yoshioka}, \bibinfo{person}{Tsutomu
  Matsumoto}, \bibinfo{person}{Takahiro Kasama}, {and}
  \bibinfo{person}{Christian Rossow}.} \bibinfo{year}{2015}\natexlab{}.
\newblock \showarticletitle{IoTPOT: analysing the rise of IoT compromises}. In
  \bibinfo{booktitle}{\emph{9th USENIX Workshop on Offensive Technologies (WOOT
  15)}}.
\newblock


\bibitem[Pnueli(1977)]%
        {ltl}
\bibfield{author}{\bibinfo{person}{Amir Pnueli}.}
  \bibinfo{year}{1977}\natexlab{}.
\newblock \showarticletitle{The temporal logic of programs}. In
  \bibinfo{booktitle}{\emph{18th Annual Symposium on Foundations of Computer
  Science (sfcs 1977)}}. IEEE.
\newblock


\bibitem[Rahmati et~al\mbox{.}(2018)]%
        {rahmati2018tyche}
\bibfield{author}{\bibinfo{person}{Amir Rahmati}, \bibinfo{person}{Earlence
  Fernandes}, \bibinfo{person}{Kevin Eykholt}, {and} \bibinfo{person}{Atul
  Prakash}.} \bibinfo{year}{2018}\natexlab{}.
\newblock \showarticletitle{Tyche: A Risk-Based Permission Model for Smart
  Homes}. In \bibinfo{booktitle}{\emph{2018 IEEE Cybersecurity Development
  (SecDev)}}. IEEE.
\newblock


\bibitem[{Raspberry Pi Foundation}(2022)]%
        {raspberrypi}
\bibfield{author}{\bibinfo{person}{{Raspberry Pi Foundation}}.}
  \bibinfo{year}{2022}\natexlab{}.
\newblock \bibinfo{booktitle}{\emph{Teach, Learn and Make with Raspberry Pi}}.
\newblock


\bibitem[Redini et~al\mbox{.}(2021)]%
        {redini2021diane}
\bibfield{author}{\bibinfo{person}{Nilo Redini}, \bibinfo{person}{Andrea
  Continella}, \bibinfo{person}{Dipanjan Das}, \bibinfo{person}{Giulio
  De~Pasquale}, \bibinfo{person}{Noah Spahn}, \bibinfo{person}{Aravind
  Machiry}, \bibinfo{person}{Antonio Bianchi}, \bibinfo{person}{Christopher
  Kruegel}, {and} \bibinfo{person}{Giovanni Vigna}.}
  \bibinfo{year}{2021}\natexlab{}.
\newblock \showarticletitle{DIANE: Identifying Fuzzing Triggers in Apps to
  Generate Under-constrained Inputs for IoT Devices}. In
  \bibinfo{booktitle}{\emph{42nd IEEE Symposium on Security and Privacy 2021}}.
\newblock


\bibitem[Ronen and Shamir(2016)]%
        {ronen2016extended}
\bibfield{author}{\bibinfo{person}{Eyal Ronen} {and} \bibinfo{person}{Adi
  Shamir}.} \bibinfo{year}{2016}\natexlab{}.
\newblock \showarticletitle{Extended functionality attacks on IoT devices: The
  case of smart lights}. In \bibinfo{booktitle}{\emph{Security and Privacy
  (EuroS\&P), 2016 IEEE European Symposium on}}. IEEE.
\newblock


\bibitem[Ronen et~al\mbox{.}(2017)]%
        {ronen2017iot}
\bibfield{author}{\bibinfo{person}{Eyal Ronen}, \bibinfo{person}{Adi Shamir},
  \bibinfo{person}{Achi-Or Weingarten}, {and} \bibinfo{person}{Colin
  O{'}Flynn}.} \bibinfo{year}{2017}\natexlab{}.
\newblock \showarticletitle{IoT goes nuclear: Creating a ZigBee chain
  reaction}. In \bibinfo{booktitle}{\emph{2017 IEEE Symposium on Security and
  Privacy (S\&P)}}.
\newblock


\bibitem[Rozier and Vardi(2007)]%
        {LTLSATMC}
\bibfield{author}{\bibinfo{person}{Kristin~Y. Rozier} {and}
  \bibinfo{person}{Moshe~Y. Vardi}.} \bibinfo{year}{2007}\natexlab{}.
\newblock \showarticletitle{LTL Satisfiability Checking}. In
  \bibinfo{booktitle}{\emph{Model Checking Software}},
  \bibfield{editor}{\bibinfo{person}{Dragan Bo{\v{s}}na{\v{c}}ki} {and}
  \bibinfo{person}{Stefan Edelkamp}} (Eds.).
\newblock
\showISBNx{978-3-540-73370-6}


\bibitem[Sha et~al\mbox{.}(2018)]%
        {sha2018iiot}
\bibfield{author}{\bibinfo{person}{Letian Sha}, \bibinfo{person}{Fu Xiao},
  \bibinfo{person}{Wei Chen}, {and} \bibinfo{person}{Jing Sun}.}
  \bibinfo{year}{2018}\natexlab{}.
\newblock \showarticletitle{IIoT-SIDefender: Detecting and defense against the
  sensitive information leakage in industry IoT}.
\newblock \bibinfo{journal}{\emph{World Wide Web}} (\bibinfo{year}{2018}).
\newblock


\bibitem[Sikder et~al\mbox{.}(2019)]%
        {sikder2019aegis}
\bibfield{author}{\bibinfo{person}{Amit~Kumar Sikder},
  \bibinfo{person}{Leonardo Babun}, \bibinfo{person}{Hidayet Aksu}, {and}
  \bibinfo{person}{A~Selcuk Uluagac}.} \bibinfo{year}{2019}\natexlab{}.
\newblock \showarticletitle{Aegis: A Context-aware Security Framework for Smart
  Home Systems}.
\newblock  (\bibinfo{year}{2019}).
\newblock


\bibitem[Tian et~al\mbox{.}(2017)]%
        {tian2017smartauth}
\bibfield{author}{\bibinfo{person}{Yuan Tian}, \bibinfo{person}{Nan Zhang},
  \bibinfo{person}{Yueh-Hsun Lin}, \bibinfo{person}{XiaoFeng Wang},
  \bibinfo{person}{Blase Ur}, \bibinfo{person}{Xianzheng Guo}, {and}
  \bibinfo{person}{Patrick Tague}.} \bibinfo{year}{2017}\natexlab{}.
\newblock \showarticletitle{SmartAuth: User-Centered Authorization for the
  Internet of Things}. In \bibinfo{booktitle}{\emph{26th {USENIX} Security
  Symposium ({USENIX} Security 17)}}.
\newblock


\bibitem[Ur et~al\mbox{.}(2013)]%
        {ur2013current}
\bibfield{author}{\bibinfo{person}{Blase Ur}, \bibinfo{person}{Jaeyeon Jung},
  {and} \bibinfo{person}{Stuart Schechter}.} \bibinfo{year}{2013}\natexlab{}.
\newblock \showarticletitle{The current state of access control for smart
  devices in homes}. In \bibinfo{booktitle}{\emph{Workshop on Home Usable
  Privacy and Security (HUPS)}}. HUPS 2014.
\newblock


\bibitem[Wang et~al\mbox{.}(2019)]%
        {AttackSurface}
\bibfield{author}{\bibinfo{person}{Qi Wang}, \bibinfo{person}{Pubali Datta},
  \bibinfo{person}{Wei Yang}, \bibinfo{person}{Si Liu}, \bibinfo{person}{Adam
  Bates}, {and} \bibinfo{person}{Carl~A. Gunter}.}
  \bibinfo{year}{2019}\natexlab{}.
\newblock \showarticletitle{Charting the Attack Surface of Trigger-Action IoT
  Platforms}. In \bibinfo{booktitle}{\emph{The 26th ACM Conference on Computer
  and Communications (CCS 19)}} (2019-11).
\newblock


\bibitem[Wang et~al\mbox{.}(2018)]%
        {wang2018fear}
\bibfield{author}{\bibinfo{person}{Qi Wang}, \bibinfo{person}{Wajih~Ul Hassan},
  \bibinfo{person}{Adam Bates}, {and} \bibinfo{person}{Carl Gunter}.}
  \bibinfo{year}{2018}\natexlab{}.
\newblock \showarticletitle{Fear and Logging in the Internet of Things}. In
  \bibinfo{booktitle}{\emph{ISOC NDSS}}.
\newblock


\bibitem[Wilson et~al\mbox{.}(2017)]%
        {wilson2017trust}
\bibfield{author}{\bibinfo{person}{Judson Wilson}, \bibinfo{person}{Riad~S
  Wahby}, \bibinfo{person}{Henry Corrigan-Gibbs}, \bibinfo{person}{Dan Boneh},
  \bibinfo{person}{Philip Levis}, {and} \bibinfo{person}{Keith Winstein}.}
  \bibinfo{year}{2017}\natexlab{}.
\newblock \showarticletitle{Trust but verify: Auditing the secure internet of
  things}. In \bibinfo{booktitle}{\emph{Proceedings of the 15th Annual
  International Conference on Mobile Systems, Applications, and Services}}.
  ACM.
\newblock


\bibitem[Yahyazadeh et~al\mbox{.}(2020)]%
        {yahyazadeh2020patriot}
\bibfield{author}{\bibinfo{person}{Moosa Yahyazadeh},
  \bibinfo{person}{Syed~Rafiul Hussain}, \bibinfo{person}{Endadul Hoque}, {and}
  \bibinfo{person}{Omar Chowdhury}.} \bibinfo{year}{2020}\natexlab{}.
\newblock \showarticletitle{PatrIoT: Policy Assisted Resilient Programmable IoT
  System}. In \bibinfo{booktitle}{\emph{International Conference on Runtime
  Verification}}. Springer.
\newblock


\bibitem[Yahyazadeh et~al\mbox{.}(2019)]%
        {yahyazadeh2019expat}
\bibfield{author}{\bibinfo{person}{Moosa Yahyazadeh}, \bibinfo{person}{Proyash
  Podder}, \bibinfo{person}{Endadul Hoque}, {and} \bibinfo{person}{Omar
  Chowdhury}.} \bibinfo{year}{2019}\natexlab{}.
\newblock \showarticletitle{EXPAT: Expectation-based policy analysis and
  enforcement for appified smart-home platforms}. In
  \bibinfo{booktitle}{\emph{Proceedings of the 24th ACM Symposium on Access
  Control Models and Technologies}}.
\newblock


\bibitem[Yu et~al\mbox{.}(2015)]%
        {yu2015handling}
\bibfield{author}{\bibinfo{person}{Tianlong Yu}, \bibinfo{person}{Vyas Sekar},
  \bibinfo{person}{Srinivasan Seshan}, \bibinfo{person}{Yuvraj Agarwal}, {and}
  \bibinfo{person}{Chenren Xu}.} \bibinfo{year}{2015}\natexlab{}.
\newblock \showarticletitle{Handling a trillion (unfixable) flaws on a billion
  devices: Rethinking network security for the internet-of-things}. In
  \bibinfo{booktitle}{\emph{ACM Workshop on Hot Topics in Networks}}.
\newblock


\bibitem[Z-Wave(2022)]%
        {zwave}
\bibfield{author}{\bibinfo{person}{Z-Wave}.} \bibinfo{year}{2022}\natexlab{}.
\newblock \bibinfo{booktitle}{\emph{Better and Safer Homes are built on
  Z-Wave}}.
\newblock Z-Wave Alliance.
\newblock


\bibitem[Zhang et~al\mbox{.}(2017)]%
        {zhang2017proximity}
\bibfield{author}{\bibinfo{person}{Jiansong Zhang}, \bibinfo{person}{Zeyu
  Wang}, \bibinfo{person}{Zhice Yang}, {and} \bibinfo{person}{Qian Zhang}.}
  \bibinfo{year}{2017}\natexlab{}.
\newblock \showarticletitle{Proximity based IoT device authentication}. In
  \bibinfo{booktitle}{\emph{IEEE INFOCOM 2017}}. IEEE.
\newblock


\bibitem[Zhang et~al\mbox{.}(2019)]%
        {autotap}
\bibfield{author}{\bibinfo{person}{Lefan Zhang}, \bibinfo{person}{Weijia He},
  \bibinfo{person}{Jesse Martinez}, \bibinfo{person}{Noah Brackenbury},
  \bibinfo{person}{Shan Lu}, {and} \bibinfo{person}{Blase Ur}.}
  \bibinfo{year}{2019}\natexlab{}.
\newblock \showarticletitle{AutoTap: Synthesizing and Repairing Trigger-Action
  Programs Using LTL Properties}. In \bibinfo{booktitle}{\emph{ICSE}}.
\newblock


\bibitem[Zhang et~al\mbox{.}(2018)]%
        {zhang2018homonit}
\bibfield{author}{\bibinfo{person}{Wei Zhang}, \bibinfo{person}{Yan Meng},
  \bibinfo{person}{Yugeng Liu}, \bibinfo{person}{Xiaokuan Zhang},
  \bibinfo{person}{Yinqian Zhang}, {and} \bibinfo{person}{Haojin Zhu}.}
  \bibinfo{year}{2018}\natexlab{}.
\newblock \showarticletitle{Homonit: Monitoring smart home apps from encrypted
  traffic}. In \bibinfo{booktitle}{\emph{Proceedings of the 2018 ACM SIGSAC
  Conference on Computer and Communications Security}}. ACM.
\newblock


\end{thebibliography}

\begin{appendix}

\section{Policy Checking}
\label{appendix:policyChecking}
\approach{}'s policy checking module conceptually takes as input 
a policy \policy, the system's execution history $\sigma$, 
and the current system state $\sigma_i$. It then checks to see whether 
the current state $\sigma_i$ satisfies all the invariants of the policy \policy. 

The main insight that \approach uses for policy checking is hinged 
upon the observation that  
any policy written in \approach{}'s policy language 
can be represented as a restricted-fragment of first-order logic formula. 
This allows us to use techniques for runtime verification for efficient 
policy checking in \approach. 

Note that, invariants in \policy can have standard past temporal 
operators such as \emph{\textbf{yesterday}} and \emph{\textbf{since}}.  
\emph{\textbf{yesterday}} is a unary temporal operator that can be applied to a linear temporal 
logic formula $\varphi$ (\ie, written \emph{\textbf{yesterday}} $\varphi$), which evaluates to 
true if and only if the formula $\varphi$ held in the immediately previous state. 
Similarly, \emph{\textbf{since}} is a binary temporal operator that can be applied to 
two linear temporal 
logic formulas $\varphi_1$ and $\varphi_2$ (\ie, written $\varphi_1$ \emph{\textbf{yesterday}} $\varphi_2$), 
which evaluates to 
true if and only if the formula $\varphi_2$ held in \emph{some} previous state, 
then $\varphi_1$ must hold in every state 
following the state in which $\varphi_2$ held until the current state. 

Since invariants in \approach can refer to arbitrary past system states through 
the use of temporal operators \emph{\textbf{yesterday}} and \emph{\textbf{since}},  
it seems natural that one has to keep track of all the previous states. 
However, as we will show, using insights from runtime verification, 
one does not need to keep track of the whole execution 
history $\sigma$ and can rather cache all previous state concisely 
into a bit-vector of size $\sigma$ ((i.e., the size of a formula is the 
number of sub-formulas in it)). This is particularly  
crucial  because $\sigma$ keeps 
growing as the system executes and would have made policy checking prohibitively expensive. 
The policy checking complexity of \approach, fortunately, is linear to the size of the 
invariant, thanks 
to its dynamic programming algorithm that is presented below.    

Suppose the temporal logic formula corresponding to a policy \policy is $\varphi$. 
One can use dynamic programming to check whether $\varphi$ 
holds in the current position $i$, denoted by $\llbracket\varphi\rrbracket^i = \mathsf{True}$. 
Instead of storing the full history of system execution $\sigma$, 
this approach only stores the values of 
$\llbracket\pi\rrbracket^{i-1}$ and $\llbracket\pi\rrbracket^{i}$, 
for all sub-formulas $\pi$ of $\varphi$. 
Thus, we need two bits of information for 
each sub-formula of $\varphi$ to check whether $\varphi$ 
holds in the current position $i$, denoted by $\llbracket\varphi\rrbracket^i = \mathsf{True}$, 
in the following way. 
\begin{align*}
\llbracket p\rrbracket^i &= p\in\sigma_i  \text{ }(p\text{ holds in } \sigma_i)\\
\llbracket \varphi_1\wedge\varphi_2\rrbracket^i &= \llbracket\varphi_1\rrbracket^i \wedge \llbracket\varphi_2\rrbracket^i\\
\llbracket \neg\varphi\rrbracket^i &=  \neg\llbracket\varphi\rrbracket^i\\
\llbracket \mathbf{yesterday}\text{ }\varphi\rrbracket^i &= i>0 \wedge   \llbracket\varphi\rrbracket^{i-1}\\
\llbracket \varphi_1\text{ }\mathbf{since}\text{ }\varphi_2\rrbracket^i &= 
\llbracket\varphi_2\rrbracket^i \vee (\llbracket\varphi_1\rrbracket^i\wedge \llbracket\varphi_1\text{ }\mathbf{since}\text{ }\varphi_2\rrbracket^{i-1})\\
\end{align*} 

As an example, $\mathbf{yesterday}\text{ }\varphi$ holds in 
the current position $i$ of the system execution history, denoted 
by $\llbracket \mathbf{yesterday}\text{ }\varphi\rrbracket^i$ if 
and only if $i>0$ and $\varphi$ held in state $i-1$, denoted 
by $\llbracket\varphi\rrbracket^{i-1}$. This is what is captured 
above in \approach{}'s policy checking algorithm presented just above. 

\section{Full Proof for \ISSP} \label{appendix:isspProof}
\begin{theorem}[Decision \ISSP]
	Given a programmable IoT system $I=\langle \cS, \cV, \cD, \cA, \policy, R\rangle$ where $\policy\in\cP$ and 
	an unsafe state $s\in\cS$, deciding whether $I$ can be transitioned to a safe state $s^\star\in\cS$ according 
	to \policy is an undecidable problem.    
\end{theorem}
 
\noindent \emph{Proof.} We reduce the halting problem to checking whether an IoT system $I$ will  be safe under $\policy$.
More specifically, our reduction will show that a given Turing machine with an input tape will halt \emph{if and only if} the encoded system will remain safe under the encoded policy.
Our proof is inspired by the undecidability proof for protection systems in operating systems by Harrison et al. \cite{harrison1976protection}

Consider a Turing machine $\mathbb{T} = \langle \mathbb{A}, \mathbb{S}, q_0, \delta \rangle$.
The Turing machine is associated with an infinite tape divided into cells, and a tape head held over a cell on the tape which can read and write on the current tape cell and move to the left-adjacent or right-adjacent cell.
$\mathbb{A}$ is the machine's alphabet of symbols with a distinct ``blank'' symbol usually denoted by \textsf{b}.
Each cell on the tape can hold one symbol.
$\mathbb{S}$ is the set of states that can be assumed by the Turing machine.
$q_0 \in \mathbb{S}$ is the initial state $\mathbb{T}$ is in.
How $\mathbb{T}$ operates is determined by the total transition function $\delta: \mathbb{A} \times \mathbb{S} :\rightarrow \mathbb{A} \times \mathbb{S} \times \{L,R\}$ i.e. how the tape head of $\mathbb{T}$ moves and how state changes happen are controlled by $\delta$.
Given that the tape head of $\mathbb{T}$ is on a tape cell containing tape symbol $x$ and $\mathbb{T}$ is in state $P$, we have $\delta(x,P) = (y,Q,R)$.
This represents the tape head writing symbol $y$ in the current tape cell, changing the state of $\mathbb{T}$ from $P$ to $Q$ and finally moving the tape head to the right-adjacent cell.
Similarly, if we had $\delta(x,P) = (y,Q,L)$ under the same circumstances, the tape head will be moved to the left-adjacent cell.
The halting problem asks if given a Turing machine $\mathbb{T} = \langle \mathbb{A}, \mathbb{S}, q_0, \delta \rangle$, started on an input
tape, will enter some predetermined state $q_f$, also called the \emph{halting} state.

We model the halting problem generally as follows.
The execution of $\mathbb{T}$ up to $i$ steps is represented by a finite trace $\sigma_i$ such that (1) $\sigma_i$ represents the current state of $\mathbb{T}$ and its tape $t$ and consequently, the current state of policy $\policy_\mathbb{T}$ generated by $\mathbb{T}$ and the state of the IoT system respectively, (2) $\sigma_i$ \emph{violates} $\policy_{\mathbb{T}}$ generated by $\mathbb{T}$
, (3) only actions that model the execution of step $i+1$ can extend $\sigma_i$ into $\sigma_{i+1}$ and (4) if $\mathbb{T}$ enters some state $q_f$, then the state of the tape in $\sigma_i$ is a state $s$ in I that satisfies $\policy_{\mathbb{T}}$.

We encode $\mathbb{T}$ as $\policy_{\mathbb{T}}$ as follows.
We represent IoT system states through the states and tape symbols of $\mathbb{T}$.
At some point in time, the Turing machine $\mathbb{T}$ has scanned some finite prefix of its tape i.e. cells $1,2,\dots,k$,
This is represented as an ordered sequence $V$ of $k$ IoT device state variables $v_1, v_2, \dots, v_k$.
Device state variable $v_i$ represents tape cell $i$ and the tape symbol $a$ being in cell $i$ represents the current value of IoT device state variable $v_i$ as $a$. 
The current Turing machine state $q$ and the current tape head being in position $j$ is represented by the current IoT system state $s$ at state $q$ and a current \emph{position} variable $p$ set to $j$, where $j \in \mathbb{N}$.
Subsequently, we describe the IoT system $I$ as an infinite set of system variables $\mathcal{X} = \{s\} \cup V \cup \{p\}$ where the domain of $s$ is the set of states $\mathbb{S}$, the domain of $V$ is the alphabet $\mathbb{A}$ and the domain of $p$ is $\mathbb{N}$.
The Turing machine operation of writing value $l$ in cell $i$ is represented as setting the value of IoT device state variable $v_i$ to $l$ i.e $v_i.setState(l)$, while the Turing machine operation of transitioning to state $q_1$ is represented by the IoT system state $s$ being set to state $q_1$ i.e. $s.setState(q_1)$.
The transition function $\delta : \mathbb{A} \times \mathbb{S} \rightarrow \mathbb{A} \times \mathbb{S} \times \{L,R\}$ is encoded as the enforcement of policy $\policy_\mathbb{T}$ through corrective actions.
To illustrate, we consider an example Turing machine transition which moves the tape head right i.e. $\delta(b,q_0) = (a,q_1,R)$, with the left case being symmetric.
Consider the Turing machine $\mathbb{T}$ to be in current state $q_0$, tape head over position $p$ on the tape, and the tape cell $p$ holding symbol $b$.
This is equivalent to the IoT system $I$ with system state $s$ at state $q_0$, position variable $p$ and IoT device variable $v_p$ set to value $b$.
The policy $\policy_\mathbb{T}$ considers the IoT system \emph{safe} when the current system state and the value of the currently observed IoT device state \emph{implies} that the current system state is a safe state i.e. $(s.state = q_0 \wedge v_p = b) \Rightarrow q_0 = q_f$.
When this does not hold, $\policy_\mathbb{T}$ prescribes a corrective action that sets the value of $v_p$ to $b$ i.e. $v_i.setState(b)$, sets the IoT system state to $q_1$ i.e. $s.setState(q_1)$, and increments the position variable i.e. $p = p+1$. 
This is reflected as the Turing machine operation of writing $b$ in the tape cell $p$, changing its state to $q_1$ and moving the tape head right to the new position $p+1$.
We can see from this description that the Turing machine $\mathbb{T}$ will halt \emph{if and only if} the encoded policy determines that the IoT system has reached a safe state i.e. $\mathbb{T}$ reaches state $q_f$.
\begin{lstlisting}[caption=An example trigger-action rule in \oh, label=code:TAP,frame=single]
    rule "Morning Routine"
        when
            Item BedSensor received update OutOfBed
        then
            WindowBlinds.sendCommand(UP)
            CoffeeMachine.sendCommand(Brew)
            Toaster.sendCommand(Start)
    end
    \end{lstlisting}
    \vspace*{-0.2in}
\end{appendix}

\end{document}